\documentclass[journal]{IEEEtran}
\usepackage{cite}
\usepackage{amsmath,amssymb,amsfonts}
\usepackage{algorithmic}

\usepackage{algorithm}

\usepackage{graphicx}
\usepackage{textcomp}
\usepackage{xcolor}
\newcommand{\RNum}[1]{\uppercase\expandafter{\romannumeral #1\relax}}
\usepackage{subfigure}

\begin{document}

\title{Enhanced Audit Bit Based Distributed Bayesian Detection in the Presence of Strategic Attacks}

\author{Chen~Quan, Baocheng~Geng, Yunghsiang~S.~Han,~\IEEEmembership{Fellow,~IEEE}    and~Pramod~K.~Varshney,~\IEEEmembership{Life~Fellow,~IEEE}
\thanks{C. Quan and P. K. Varshney are with the Department of Electrical Engineering and Computer
Science, Syracuse University, Syracuse, NY 13244 USA (e-mail: \{chquan,varshney\}@syr.edu).}
\thanks{B. Geng was with the Department of Electrical Engineering and Computer Science, Syracuse University, Syracuse, NY 13244 USA. He is now with the Department of Computer
Science, University of Alabama at Birmingham, Birmingham, AL 35205 USA.  (e-mail: bgeng@uab.edu).}
\thanks{Y. S. Han is with the Shenzhen Institute for Advanced Study, University of Electronic Science and Technology of China, Shenzhen, China
(e-mail: yunghsiangh@gmail.com).}
}
\maketitle

\maketitle

\IEEEpeerreviewmaketitle
\begin{abstract}
	This paper employs an audit bit based mechanism to mitigate the effect of Byzantine attacks. In this framework, the optimal attacking strategy for intelligent attackers is investigated for the traditional audit bit based scheme (TAS) to evaluate the robustness of the system. We show that it is possible for an intelligent attacker to degrade the performance of TAS to the system without audit bits. To enhance the robustness of the system in the presence of intelligent attackers, we propose an enhanced audit bit based scheme (EAS). The optimal fusion rule for the proposed scheme is derived and the detection performance of the system is evaluated via the probability of error for the system. Simulation results show that the proposed EAS improves the robustness and the detection performance of the system. Moreover, based on EAS, another new scheme called the reduced audit bit based scheme (RAS) is proposed which further improves system performance. We derive the new optimal fusion rule and the simulation results show that RAS outperforms EAS and TAS in terms of both robustness and detection performance of the system. Then, we extend the proposed RAS for a wide-area cluster based distributed wireless sensor networks (CWSNs). Simulation results show that the proposed RAS significantly reduces the communication overhead between the sensors and the FC, which prolongs the lifetime of the network.
\end{abstract}
\section{Introduction}
Distributed detection in wireless sensor networks (WSNs) has been studied over the last few decades \cite{varshney2012distributed}\cite{veeravalli2012distributed}. In distributed WSNs, instead of sending raw observations, the sensors send their quantized observations or their hard/soft decisions regarding the presence or absence of the phenomenon of interest (PoI) to the fusion center (FC) to make the final decision. This distributed framework is attractive for sensor networks that employ battery-limited sensors in bandwidth-limited environments. Because of the advantages of the distributed mechanism, it is widely used in many applications, such as IoT, cognitive radio networks, object detection networks, distributed spectrum sensing and military surveillance systems.

Security is an important issue for the distributed WSNs. The openness of the wireless networks and the distributed nature of such networks make the distributed system more vulnerable to various attacks. The security issues associated with distributed networks are increasingly being studied, e.g., jamming, wiretap, spoofing \cite{jover2014enhancing}\cite{gai2017spoofing}
and Byzantine attacks\cite{zhang2015byzantine}\cite{lamport2019byzantine}. In this paper, we focus on Byzantine attacks. When the system suffers from Byzantine attacks, some sensors in the network might be compromised and fully controlled by intelligent adversaries. We refer to these compromised sensors as Byzantine nodes. They may send falsified information to the FC. There are several types of Byzantine attacks, such as independent probabilistic attack\cite{penna2011detecting}, dependent probabilistic attack\cite{kailkhura2013optimal} and non-probabilistic attack\cite{wang2014secure}. In probabilistic attacks, the Byzantine nodes are in pursuit of long-term profits by launching attacks with a certain probability. In non-probabilistic attacks,
the Byzantine nodes decide to launch attacks only when the observations satisfy some specific conditions. For example, a Byzantine node decides to launch attacks only when its observations are higher than threshold $\lambda_1$ or lower than threshold $\lambda_0$, where $\lambda_1>\lambda_0$. 

There are several works that have studied Byzantine attack issues in distributed detection systems. In \cite{kailkhura2013covert},optimal intelligent data falsification attacks on distributed detection systems are studied. The smart attackers attempt to constrain their exposure to the defense mechanism and maximize the attacking efficacy. 
In \cite{vempaty2014false}, an adaptive algorithm at the FC is proposed to mitigate the impact of Byzantine attacks in the false discovery rate based distributed detection system when the Byzantine nodes know the true hypothesis. 
In \cite{li2010catch}, \cite{rawat2010collaborative}, distributed detection problems are investigated in the context of collaborative spectrum sensing under Byzantine attacks. An abnormality-detection-based algorithm for the detection of attackers in collaborative spectrum sensing is proposed\cite{li2010catch}. In \cite{rawat2010collaborative}, the condition under which the Byzantine attackers totally blind the FC is investigated and an algorithm is proposed to detect Byzantine attacks by counting the mismatches between the local decisions and the global decision at the FC. In \cite{kailkhura2015distributed}, the optimal attacking strategies are analyzed in the distributed network for the cases where the FC has the knowledge of the attackers' strategy and where the FC does not know the attackers' strategy. Audit based mechanisms are proposed to mitigate the effect of Byzantine attacks on the distriFbuted WSNs\cite{hashlamoun2017mitigation}\cite{hashlamoun2018audit}. In \cite{hashlamoun2017mitigation}, the audit bit based distributed detection scheme is proposed in the Neyman-Pearson  framework by utilizing Kullback-Leibler divergence (KLD) to characterize the detection performance of the system. Each sensor sends one additional audit bit to the FC which gives some information about the behavioral identity of each sensor and improves the detection and security performance of the system. Improved system robustness to Byzantine attacks is achieved at the expense of increased communication overhead. In \cite{hashlamoun2018audit}, the audit bit based mechanism is utilized in the Bayesian setting. The detection performance of the system is evaluated in terms of the probability of error of the global decision at the FC, and the mitigation scheme over time is proposed by using the information coming from the audit bits.

Our work is most related to the works in \cite{hashlamoun2017mitigation} and \cite{hashlamoun2018audit}. In \cite{hashlamoun2017mitigation} and \cite{hashlamoun2018audit}, all the sensors in the network are divided into groups of two. Each sensor sends its local decisions to the FC via two paths, one is direct path and another is through the sensor in the same group (indirect path). The indirect decision bits that reach the FC via indirect path are referred to as audit bits which gives us extra information about the behavioral identity of each sensor. In [15] and [16], it is assumed that each Byzantine node falsifies its own local decisions and the decisions coming from its group member with the same probability. However, different from the existing works in \cite{hashlamoun2017mitigation} and \cite{hashlamoun2018audit}, we consider a more realistic case in which the strong assumption of Byzantine nodes' attack behavior made in \cite{hashlamoun2017mitigation} and \cite{hashlamoun2018audit}, namely of equal probability, is relaxed. We call this type of Byzantine nodes as intelligent attackers. We show that the traditional audit bit based scheme (TAS) is not robust enough in the presence of intelligent attackers. Two new schemes, which are the enhanced audit bit based scheme (EAS) and the reduced audit bit based scheme (RAS), are proposed to improve the robustness and the detection performance of the system under intelligent attacks. Then, we extend the above RAS for cluster based wide-area wireless sensor networks (CWSNs) \cite{lindsey2002pegasis}\cite{manjeshwar2001teen}. The cluster based framework has been proposed to deal with the significantly increased energy consumption of the sensors due to the long distance transmission in wide-area networks \cite{masazade2012dynamic}\cite{niu2005distributed}. This framework not only ensures higher data transmission efficiency, larger network scale, lower bandwidth consumption and prolonged network lifetime, but also efficiently reduces the amount of information transmission in the entire network and mitigates energy dissipation due to collisions. In CWSNs, sensors are divided into several clusters and each cluster is equipped with one cluster head (CH) which has ample energy and computation capacities for operation purposes. The CHs are responsible for collecting the data in the cluster and sending it to the FC. In this work, the sensors in each cluster are further divided into groups of two. Each sensor sends its own decisions via direct and indirect path to the corresponding CH just like the previously proposed audit-based system \cite{hashlamoun2017mitigation} and \cite{hashlamoun2018audit}. The data aggregation rule for the CHs are designed according to RAS which prolongs the lifetime of the networks with the improved detection performance of the system.\footnote{This framework is also suitable for sensor networks with mobile access points (SENMA) where the CHs traverse the network to collect information directly from the sensors \cite{tong2003sensor}.} We assume that CHs have ample energy to support the long distance transmission\footnote{The CHs are assumed to be small base stations that can be charged or be unmanned aerial vehicles (UAVs) that are equipped with energy harvesting (EH) circuits which enable the CHs to harvest energy from renewable sources, e.g., vibration, solar and
wind, to replenish their energy buffers \cite{sudevalayam2010energy}.} and some protections against the attacks so that they can be trusted by the FC, e.g., tamper-resistant security module \cite{walters2007wireless}\cite{perrig2004security}. The main contributions of this work are summarized as follows:
\begin{itemize}
	\item We derive the detection performance of the system that employs TAS in the presence of intelligent attackers. Instead of considering an identical attacking strategy in which each sensor utilizes the same attacking probability to falsify its own decisions and the decisions coming from their group member \cite{hashlamoun2017mitigation}\cite{hashlamoun2018audit}, we consider attackers that can use different attacking strategies. The optimal attacking strategy of intelligent attackers is investigated and we show that it is possible to degrade the performance of TAS to the system without audit bits.
	\item An EAS is proposed to deal with the security issues arising from the intelligent attackers that may use different attacking strategies. We derive the optimal decision rule at the FC and evaluate its detection performance. Simulation results show that the proposed EAS outperforms TAS and the direct scheme under both intelligent attacks and non-intelligent attacks. 
	\item The scheme EAS is further extended and a new scheme namely RAS is proposed based on our newly proposed EAS. We show that RAS is able to further improve the robustness and the detection performance of the system. 
	\item A wide-area cluster-based WSN is considered. We extend the proposed RAS and design the data aggeration rule for the CHs. Simulation results show a significant reduction in the overall communication overhead between the FC and the CHs.
\end{itemize}

The rest of the paper is organized as follows. Section \RNum{2} presents the system model of TAS. The optimal attacking strategy is investigated for intelligent attackers and the detection performance of the system is evaluated under intelligent attacks. Section \RNum{3} presents the proposed EAS and evaluates the detection performance and the robustness of the system. Section \RNum{4} presents the proposed RAS and extends it for the wide-area networks with several clusters. Section \RNum{5} presents some concluding remarks.
\section{Traditional Audit Bit Based Scheme Under Intelligent Attacks}
We consider the binary hypothesis testing problem assuming that there are two possible hypotheses, ${H}_0$ (signal is absent) and ${H}_1$ (signal is present), regarding a phenomenon of interest (PoI). Consider that we deploy a cluster
of $N$ sensors to determine which of the two hypotheses is true. Based on the local observations, each sensor $i\in\{1,\dots,N\}$ makes a binary decision $v_i\in\{0,1\}$ regarding the true hypothesis using the likelihood ratio (LR) test
\begin{equation}
    \frac{P(y_i|\mathcal{H}_1)}{P(y_i|\mathcal{H}_0)}\overset{v_i=1}{\underset{v_i=0}{\gtrless}}\lambda,
\end{equation}where $\lambda$ is the identical threshold used by all the sensors \cite{tsitsiklis1988decentralized}, and, $P(y_i|\mathcal{H}_m)$ denotes the conditional probability density function (PDF) of observation $y_i$ under the hypothesis $\mathcal{H}_m$, for $m=0,1$. In the audit bit based framework \cite{hashlamoun2017mitigation} \cite{hashlamoun2018audit}, the $N$ sensors are partitioned into $G$ groups where each group $g\in \{1,\dots,G\}$ is composed of two sensors.\footnote{The sensors are divided into groups of two based on certain criteria, e.g., according to their distances from each other.} Let $i$ and $j$ represent the sensors in the same group, where $i\in\{1,2,\dots,N\}$ and $j\in\{1,2,\dots,i-1,i+1,\dots,N\}$. Each sensor $i$ sends its local binary decision to the FC via two paths, one is direct and the other is through sensor $j$ in the same group. At the FC, we design a match  and  mismatch detector  (MMSD)  module that detects if the sensor's direct decision matches or mismatches the decision sent through sensor $j$ (indirect decision). 

\begin{figure}[htb]
  \centering
  \subfigure[The architecture of group k.]{
    \includegraphics[width=12em,height=11em]{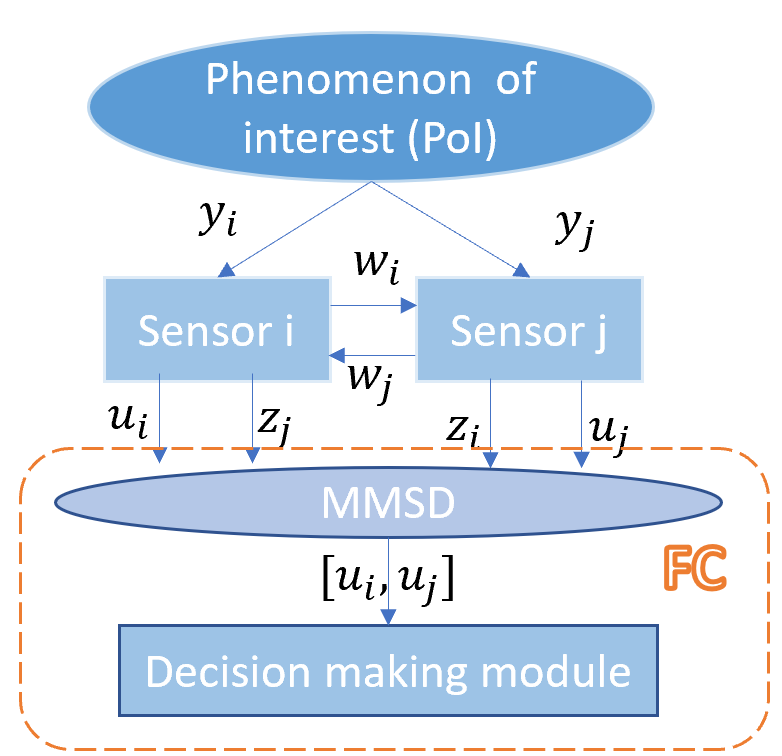}}
      \subfigure[The overall detection network for audit bit based scheme.]{
    \includegraphics[width=15em,height=11em]{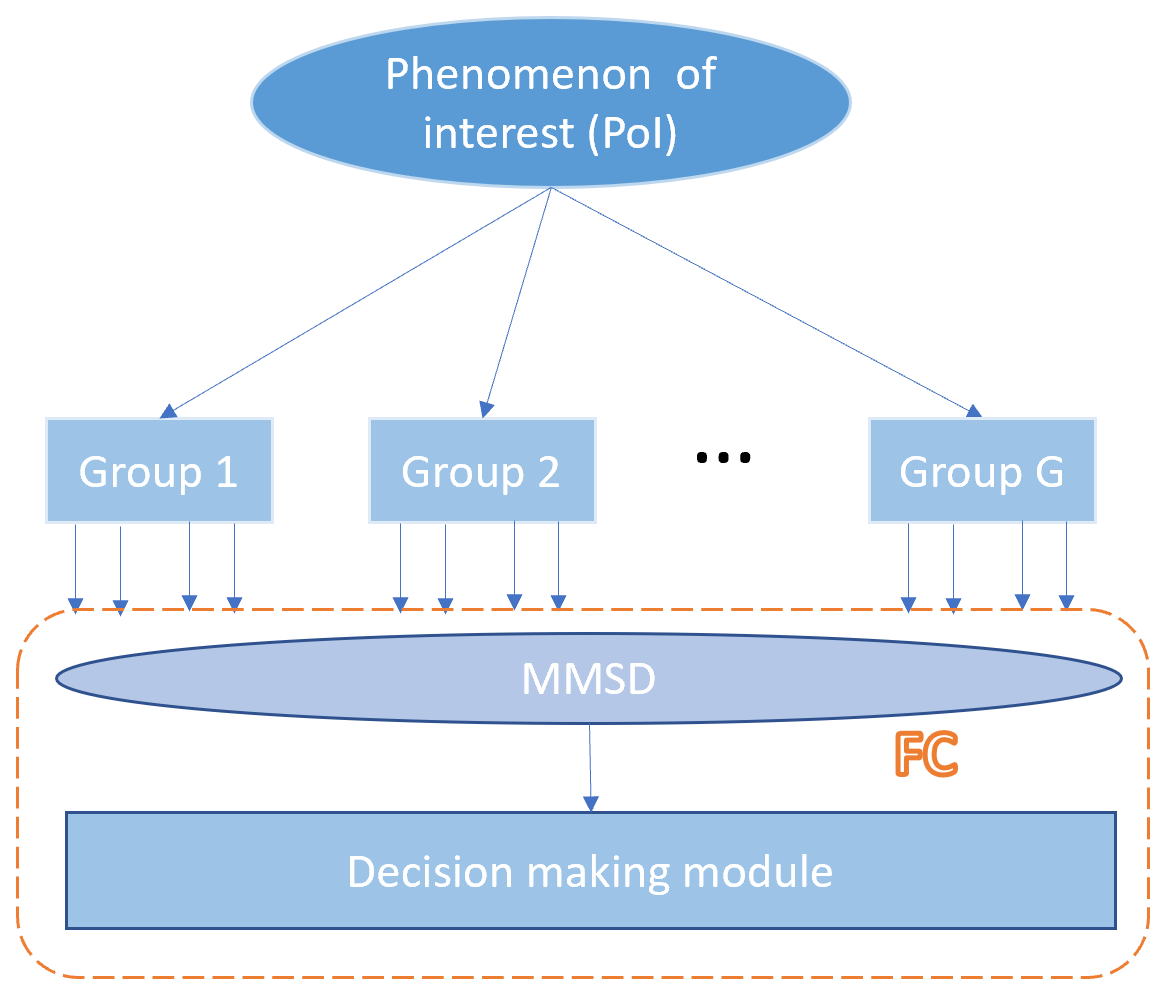}}
  \caption{The architecture of any group $k$ is shown in (a) and the overall system model is shown in (b).}
  \label{Fig.main} 
\end{figure}

The architecture of each group is shown as Fig.~\ref{Fig.main}(a) and the overall detection network for TAS is shown as Fig.~\ref{Fig.main}(b).
As shown in Fig.~\ref{Fig.main}(a), after making its own decision $v_i$, sensor $i$ sends (i) $u_{i}$ directly to the MMSD; (ii) $w_{i}$ to the sensor $j$ in the same group; (iii) $z_{j}$, corresponding to $w_j$ coming from the sensor $j$ in the same group, to the MMSD. Similarly, sensor $j$ also sends two decisions $u_{j}$ and $z_{i}$ to the MMSD.
If the sensor $i$ is a Byzantine node, i.e., $i= B$, the decisions $v_i, w_i$and $u_i$ are not necessarily the same and $z_j$ are also not necessarily equal to $u_j$. Let $p(v_i\neq u_i|i=B)$, $p(v_i\neq w_i|i=B)$ and $p(w_j\neq z_j|i=B)$ denote the probabilities that the Byzantine node $i$ flips its own decision, flips the decision sent to its group member and flips the decision coming from its group member, respectively. The probabilities  $p_2=p(w_j\neq z_j|i=B)$ and $p_1=p(v_i\neq u_i|i=B)=p(v_i\neq w_i|i=B)$ are the attacking parameters the attackers want to optimize. If the sensor $i$ is honest, i.e., $i= H$, we have $v_i=w_i=u_i$ and $z_j=u_j$. In other words, $p(v_i\neq u_i|i=H)=p(v_i\neq w_i|i=H)=0$. We assume that a fraction $\alpha_0$ of the $N$ sensors are Byzantine nodes and the FC is not aware of the identity of Byzantine nodes in the network. Hence, each node has the probability of $\alpha_0$ to be a Byzantine node. We also assume that each Byzantine node attacks the network independently with a certain probability.

After collecting all the local decisions, the MMSD makes binary decisions regarding the match and mismatch (MMS) status of the two decisions corresponding to the same sensor received over different paths, i.e., whether or not the decisions sent via different paths are the same, for all the sensors. Let $d_i$ represent the MMS status of sensor $i$ which is called the status indicator of sensor $i$. To give a concrete illustration, take one group of sensors $(i,j)$ as an example. The MMSD sets $d_j=1$ when $u_i=z_i$ and $d_j=0$ when $u_i\neq z_i$. Similarly, the MMSD sets $d_i=1$ if $u_j=z_j$ and $d_i=0$ if $u_j\neq z_j$. The decisions $d_i$ and $d_j$ are the status indicators of sensor $i$ and sensor $j$, respectively. According to the status indicator for each sensor, the FC places the sensors into two sets $\underline{S}$ and $\overline{S}$. Set $\underline{S}$ contains the sensors whose status indicators are equal to 1 and Set $\overline{S}$ contains the sensors whose status indicators are equal to 0. By employing the extra information coming from these status indicators, we are able to improve the detection performance of the system.

In the following two subsections, we discuss two different attack models and investigate the robustness of the traditional audit bit based mechanism under these two types of attacks. One attack model{\footnote{This attack model follows the work in \cite{hashlamoun2017mitigation} and \cite{hashlamoun2018audit}.}} is that the Byzantine nodes are assumed to flip their own decisions and all the decisions they received with the same probability $p$, i.e., $p_1=p_2=p$. The other model is that the Byzantine nodes use different probabilities to flip their own decisions and all the decisions they receive, i.e., $p_1\neq p_2$. It is more general and practical to consider Byzantine nodes which relax the assumption of $p_1=p_2=p$ made in the traditional audit bit based mechanism. This allows the Byzantines to be intelligent by optimally employing unequal probabilities $p_1$ and $p_2$. 

\subsection{TAS}
 In the traditional audit bit based mechanism, the Byzantine nodes are assumed to flip their own decisions and all the decisions they receive with the same probability $p$, i.e., $p_1=p_2=p$. Based on the status indicators $\{d_i\}_{i=1}^N$, we have the following two cases \cite{hashlamoun2017mitigation}. 
\paragraph{If $d_i=1$} $i$ is a Byzantine node with probability

\begin{equation}
\begin{split}
    \underline{\alpha}&=P(i=B|d_i=1)\\
    &=\frac{\alpha_0(1-p)[1-2\alpha_0p(1-2p)]}{1-\alpha_0(3-2p)p+4\alpha_0^2(1-p)p^2}
\end{split}
\end{equation}
and the sensor $i$ is placed in set $\underline{S}$.
\paragraph{If $d_i=0$} $i$ is a Byzantine node with probability

\begin{equation}
\begin{split}
    \overline{\alpha}&=P(i=B|d_i=0)\\
    &=\frac{1+2(1-p)(\alpha_0-2\alpha_0p))}{1+2(1-p)(1-2\alpha_0p))}
\end{split}
\end{equation}
and the sensor $i$ is placed in set $\overline{S}$.

It has been proved in \cite{hashlamoun2018audit} (Lemma 1) that $\underline{\alpha}\leq \alpha_0 \leq \overline{\alpha}$.
In other words, all the sensors are divided into two sets $\underline{S}$ and $\overline{S}$ in which the sensors have lower probability $\underline{\alpha}$ and higher probability $\overline{\alpha}$ of being Byzantine nodes, respectively, according to status indicators $\mathbf{d}=[d_1,d_2,\dots,d_N]$.  
Let $P_d$, $P_f$ be the probability of detection and the probability of false alarm for any sensor $i\in\{1,\ldots, N\}$, respectively, i.e., $P_d=P(v_i=1|\mathcal{H}_1)$ and $P_f=P(v_i=1|\mathcal{H}_0)$. Thus, the probability mass function (pmf) of local decision $u_i$ is expressed as
\begin{equation}\label{eq:P_ui}
    P(u_i|\mathcal{H}_q)=
\begin{cases}
\underline{\pi}_{1q}^{u_i}(1-\underline{\pi}_{1q})^{1-{u_i}}& \text{for $i\in\underline{S}$}\\
\overline{\pi}_{1q}^{u_i}(1-\overline{\pi}_{1q})^{1-{u_i}}& \text{for $i\in\overline{S}$}
\end{cases}
\end{equation}
for q=0,1, where, for $i\in \underline{S}$,
\begin{subequations}\label{eq:pi_11_10_under}
\begin{align}
    &\underline{\pi}_{11}=1-\underline{\pi}_{01}=P(u_{i}=1|\mathcal{H}_1)=P_d(1-\underline{\alpha}p)+\underline{\alpha}p(1-P_d)\\
    &\underline{\pi}_{10}=1-\underline{\pi}_{00}=P(u_{i}=1|\mathcal{H}_0)=P_f(1-\underline{\alpha}p)+\underline{\alpha}p(1-P_f)
\end{align}
\end{subequations}
and, for $i\in \overline{S}$,
\begin{subequations}\label{eq:pi_11_10_upper}
\begin{align}
    &\overline{\pi}_{11}=1-\overline{\pi}_{01}=P(u_{i}=1|\mathcal{H}_1)=P_d(1-\overline{\alpha}p)+\overline{\alpha}p(1-P_d)\\
    &\overline{\pi}_{10}=1-\overline{\pi}_{00}=P(u_{i}=1|\mathcal{H}_0)=P_f(1-\overline{\alpha}p)+\overline{\alpha}p(1-P_f).
\end{align}
\end{subequations}
Then the optimal decision rule when the attacking strategy $p$ is assumed to be known is given as
\begin{equation}\label{eq:LLR_simple}
    \underline{W}\underline{U}+ \overline{W}\overline{U}\gtrless \eta^{(A)},
\end{equation}
where $\underline{U}=\sum_{i\in\underline{S}}u_i$, $\overline{U}=\sum_{i\in\overline{S}}u_i$, $\underline{W}=\log(\frac{\underline{\pi}_{11}(1-\underline{\pi}_{10})}{\underline{\pi}_{10}(1-\underline{\pi}_{11})})$,  $\overline{W}=\log(\frac{\overline{\pi}_{11}(1-\overline{\pi}_{10})}{\overline{\pi}_{10}(1-\overline{\pi}_{11})})$, $\eta^{(A)}=\log(\frac{\pi_0}{\pi_1})+\underline{N}\log(\frac{1-\underline{\pi}_{10}}{1-\underline{\pi}_{11}})+\overline{N}\log(\frac{1-\overline{\pi}_{10}}{1-\overline{\pi}_{11}})$, $\underline{N}=|\underline{S}|$, and $\overline{N}=|\overline{S}|$. Note that $\underline{U}$ and $\overline{U}$ are binomial distributed random variables with parameters $(N,\underline{\pi}_{10})$ and $(N,\overline{\pi}_{10})$, respectively, under $\mathcal{H}_0$, and with parameters $(N,\underline{\pi}_{11})$ and $(N,\overline{\pi}_{11})$, respectively, under $\mathcal{H}_1$. When $N$ is large, $\underline{N}$ and $\overline{N}$ can be approximated by their expected value $NP(u_i=z_i)$ and $NP(u_i\neq z_i)$. $\eta^{(A)}$ is the threshold used by the FC for the traditional audit bit based system, where $\eta^{(A)}=\log(\frac{\pi_0}{\pi_1})+NP(u_i=z_i)\log(\frac{1-\underline{\pi}_{10}}{1-\underline{\pi}_{11}})+NP(u_i\neq z_i)\log(\frac{1-\overline{\pi}_{10}}{1-\overline{\pi}_{11}})$. Moreover, $\underline{U}$ and $\overline{U}$ can be approximated by the Gaussian distribution with parameters given as follows:
\begin{subequations}\label{eq:mu_sigma}
    \begin{align}
        \mu_0^{(A)}=&E[U|\mathcal{H}_0]\notag\\
        =&N[P(u_i=z_i)\underline{\pi}_{10}\underline{W}+P(u_i\neq z_i)\overline{\pi}_{10}\overline{W}]\\
        \mu_1^{(A)}=&E[U|\mathcal{H}_1]\notag\\=&N[P(u_i=z_i)\underline{\pi}_{11}\underline{W}+P(u_i\neq z_i)\overline{\pi}_{11}\overline{W}]\\
        (\sigma_0^{(A)})^2=&Var[U|\mathcal{H}_0]=\underline{N}[P(u_i=z_i)\underline{\pi}_{10}(1-\underline{\pi}_{10})\underline{W}^2\notag\\
        &+P(u_i\neq z_i)\overline{\pi}_{10}(1-\overline{\pi}_{10})\overline{W}^2]\\
        (\sigma_1^{(A)})^2=&Var[U|\mathcal{H}_0]=\underline{N}[P(u_i=z_i)\underline{\pi}_{11}(1-\underline{\pi}_{11})\underline{W}^2\notag\\
       &+P(u_i\neq z_i)\overline{\pi}_{11}(1-\overline{\pi}_{11})\overline{W}^2].
    \end{align}
\end{subequations}

The detection performance, characterized by the probability of error $P_e^{(A)}$ for the system with TAS, is given as
\begin{equation}\label{eq:pe}
    P_e^{(A)}=\pi_0Q\left(\gamma_f^{(A)}\right)+\pi_1Q\left(\gamma_m^{(A)}\right),
\end{equation}
where $\gamma_f^{(A)}=\frac{\eta^{(A)}-\mu_0^{(A)}}{\sigma_0^{(A)}}$ and $\gamma_m^{(A)}=\frac{\mu_1^{(A)}-\eta^{(A)}}{\sigma_1^{(A)}}$. Let $P_e^{(D)}$ denote the probability of error for the system with direct scheme, which is expressed as \eqref{eq:pe_d}. It has been shown in \cite{hashlamoun2018audit} (Theorem 3) that the probability of error of the traditional audit based system given any $\alpha_0$ and $p$ is always less than or equal to that of the system which relies only on direct decisions, i.e, $P_e^{(A)}\leq P_e^{(D)}$. 

However, due to the strong assumption of $p_1=p_2=p$, TAS can accurately assess the behavioral identity of each sensor in the network so that it can improve the detection and security performances of the system. It is obvious that a higher $p$ means a higher probability that the Byzantine nodes flip their own decisions and the decisions coming from their group members. Thus, the Byzantine nodes have a higher probability of being placed in the Set $\overline{S}$. In the next subsection, we relax the the assumption of $p_1=p_2=p$ and investigate the detection performance of the traditional audit bit based system under the relaxed assumption.

\subsection{The Intelligent Attacker under Traditional Audit Bit Based System}
To make the model more general, we assume that the attackers are more intelligent in that they can employ different values of $p_1$ and $p_2$ that are not necessarily equal. In this subsection, we analyze the detection performance of the traditional audit bit based system under such intelligent attacks. 

When the FC under intelligent attacks makes use of the status indicators to place all the sensors into two sets, we have the following two cases.
\paragraph{If $d_{i}=1$} $i$ is a Byzantine node with probability

\begin{equation}
\begin{split}
    \underline{\alpha}^{I}&=P(i=B|d_{i}=1)\\
    &=\frac{P(d_{i}=1|i=B)P(i=B)}{P(d_{i}=1)},
\end{split}
\end{equation}
where
\begin{equation}
\begin{split}
&P(d_{i}=1|i=B)=P(u_{j}=z_{j}|i=B)\\
&=\alpha_0p_1^2(1-p_2)+\alpha_0p_1(1-p_1)p_2+\alpha_0(1-p_1)^2(1-p_2)\\
&+(1-\alpha_0)(1-p_2)+\alpha_0(1-p_1)p_1p_2\\
&=-4\alpha_0 p_1^2 p_2+4\alpha_0 p_1p_2-2\alpha_0 p_1+2\alpha_0 p_1^2-p_2+1
\end{split}
\end{equation}
and
\begin{equation}
\begin{split}
&P(d_{i}=1|i=H)=P(u_{j}=z_{j}|i=H)\\
&=\alpha_0p_1^2+\alpha_0(1-p_1)^2+(1-\alpha_0)\\
&=2\alpha_0 p_1^2-2\alpha_0p_1+1.
\end{split}
\end{equation}
Thus, the unconditional probability of matching $p(u_{j}=z_{j})$ is given as
\begin{equation}
\begin{split}
    &P(d_{i}=1)\\
    &=P(u_{j}=z_{j}|i=H)P(i=H)+P(u_{j}=z_{j}|i=B)P(i=B)\\
    &=-4\alpha_0^2 p_1^2 p_2+4\alpha_0^2 p_1p_2+2\alpha_0 p_1^2-\alpha_0 p_2-2\alpha_0 p_1+1.
\end{split}
\end{equation}
In this case, the sensor $i$ is placed in set $\underline{S}$ with
\begin{equation}\label{alpha_I_1}
    \underline{\alpha}^{I}=\frac{4\alpha_0^2 p_1^2 p_2+4\alpha_0^2 p_1p_2-2\alpha_0^2 p_1+2\alpha_0^2 p_1^2-\alpha_0p_2+\alpha_0}{4\alpha_0^2 p_1^2 p_2+4\alpha_0^2 p_1p_2+2\alpha_0 p_1^2-\alpha_0 p_2-2\alpha_0 p_1+1}.
\end{equation}
\paragraph{If $d_{i}=0$} $i$ is a Byzantine node with probability

\begin{equation}\label{alpha_I_2}
\begin{split}
    \overline{\alpha}^{I}&=P(i=B|d_{i}=0)\\
    &=P(i=B|u_{j}\neq z_{j})\\
    &=\frac{P(u_{j}\neq z_{j}|i=B)P(i=B)}{P(u_{j}\neq z_{j})}\\
    &=\frac{4\alpha_0 p_1^2 p_2-4\alpha_0 p_1p_2+2\alpha_0 p_1-2\alpha_0 p_1^2+p_2}{4\alpha_0 p_1^2 p_2-4\alpha_0 p_1p_2-2 p_1^2+p_2+2 p_1},
\end{split}
\end{equation}
where $p(u_{j}\neq z_{j}|i=B)=1-p(u_{j}=z_{j}|i=B)$ and $p(u_{j}\neq z_{j})=1-p(u_{j}=z_{j})$. In this case, the sensor $i$ is placed in set $\overline{S}$. We show two important properties of $\overline{\alpha}^{I}$ and $\underline{\alpha}^{I}$ in the next lemma.
\paragraph*{Lemma 1} We have the following two relationships in terms of $\underline{\alpha}^{I}$, $\overline{\alpha}^{I}$, and $\alpha_0$.
\begin{enumerate}
	\item Under intelligent attacks, the probability of being a Byzantine node given the sensor in Set $\underline{S}$ is smaller than or equal to the one given the sensor in Set $\overline{S}$, i.e., $\underline{\alpha}^{I}\leq\alpha_0\leq \overline{\alpha}^{I}$.
	\item $\underline{\alpha}^{I}=\overline{\alpha}^{I}=\alpha_0$ when $p_2=0$.
\end{enumerate}

\begin{IEEEproof}
 According to \eqref{alpha_I_1} and \eqref{alpha_I_2}, we show that $\frac{\partial \underline{\alpha}^{I}}{\mathrm{d}p_2}\leq 0$, and $\frac{\partial \underline{\alpha}^{I}}{\mathrm{d}p_1}\leq 0$. Due to the fact that $\alpha_0\in[0,1]$, $p_1\in[0,1]$, and $p_2\in[0,1]$, we have
\begin{subequations}\label{eq:proof1}
	\begin{align}
		\frac{\partial\underline{\alpha}^{I}}{\mathrm{d}p_2}&=\frac{(4\alpha_0^2 p_1(1-p_1)-\alpha_0)(1-\alpha_0)(2\alpha_0p_1(p_1-1)+1)}{(4\alpha_0^2 p_1^2 p_2+4\alpha_0^2 p_1p_2+2\alpha_0 p_1^2-\alpha_0 p_2-2\alpha_0 p_1+1)^2}\notag\\
		&\overset{(a)}{\leq} \frac{\alpha_0(\alpha_0-1)(1-\alpha_0)(2\alpha_0p_1(p_1-1)+1)}{(4\alpha_0^2 p_1^2 p_2+4\alpha_0^2 p_1p_2+2\alpha_0 p_1^2-\alpha_0 p_2-2\alpha_0 p_1+1)^2}\notag\\
		&\overset{(b)}{\leq} 0\\
		\frac{\partial\underline{\alpha}^{I}}{\mathrm{d}p_1}&=-2\alpha_0^2(1-\alpha_0p_2)(1-2p_2)^2\leq 0.	
	\end{align}
\end{subequations}
The equality in (a) is achieved when $p_1=\frac{1}{2}$. (b) is due to the fact that $2\alpha_0p_1(p_1-1)+1\geq 1-\frac{\alpha_0}{2}> 0$ and the equality in (b) is achieved when $\alpha_0=1$. Thus, according to \eqref{eq:proof1}, we have the maximum value of $\underline{\alpha}^{I}$ when $p_1=0$ and $p_2=0$, i.e., $\underline{\alpha}^{I}(p_1,p_2)\leq \underline{\alpha}^{I}(p_1=0,p_2=0)=\alpha_0$. Since $p(u_{i}=z_{i})\underline{\alpha}^{I}+p(u_{i}\neq z_{i})\overline{\alpha}^{I}=\alpha_0$, we have
\begin{equation}
	\begin{split}
	P(u_{i}=z_{i})\alpha_0+P(u_{i}\neq z_{i})\overline{\alpha}^{I}&\geq \alpha_0\\
	P(u_{i}\neq z_{i})\overline{\alpha}^{I}&\ge\alpha_0(1-P(u_{i}= z_{i}))\\
	\overline{\alpha}^{I}&\geq \alpha_0
	\end{split}
\end{equation}
Based on the analysis above, we conclude that $\underline{\alpha}^{I}\leq\alpha_0\leq \overline{\alpha}^{I}$. Note that the equality on both sides can be achieved when $p_2=0$.
Hence, we get the results stated in Lemma 1.
\end{IEEEproof}
Substituting $\underline{\alpha}$ and $\overline{\alpha}$ with $\underline{\alpha}^{I}$ and $\overline{\alpha}^{I}$, respectively, in \eqref{eq:pi_11_10_under} and \eqref{eq:pi_11_10_upper}, we can obtain $\underline{\pi}_{10}^I$,  $\underline{\pi}_{11}^I$, $\overline{\pi}_{10}^I$,  $\overline{\pi}_{11}^I$. After getting $\underline{\pi}_{10}^I$,  $\underline{\pi}_{11}^I$ and $\overline{\pi}_{10}^I$, $\overline{\pi}_{11}^I$, we can calculate the pmfs of $u_i$ according to \eqref{eq:P_ui}. Hence, the probability of error for the system under intelligent attack is given by $P_e^I=\pi_0Q\left(\gamma_f^{I}\right)+\pi_1Q\left(\gamma_m^{I}\right)$. $\gamma_f^{I}$ and $\gamma_m^{I}$ are shown in \eqref{eq:gama_f_m}, where $D_0(\underline{\alpha}^{I},p_1,p_2)=\underline{\pi}_{10}^{(I)}\log(\frac{\underline{\pi}_{10}^{(I)}}{\underline{\pi}_{11}^{(I)}})+(1-\underline{\pi}_{10}^{(I)})\log(\frac{1-\underline{\pi}_{10}^{(I)}}{1-\underline{\pi}_{11}^{(I)}})$, $D_0(\overline{\alpha}^{I},p_1,p_2)=\overline{\pi}_{10}\log(\frac{\overline{\pi}_{10}^{(I)}}{\overline{\pi}_{11}^{(I)}})+(1-\overline{\pi}_{10}^{(I)})\log(\frac{1-\overline{\pi}_{10}^{(I)}}{1-\overline{\pi}_{11}^{(I)}})$, $D_1(\overline{\alpha}^{I},p_1,p_2)=\overline{\pi}_{11}^{(I)}\log(\frac{\underline{\pi}_{11}^{(I)}}{\overline{\pi}_{10}^{(I)}})+(1-\overline{\pi}_{11}^{(I)})\log(\frac{1-\overline{\pi}_{11}^{(I)}}{1-\overline{\pi}_{10}^{(I)}})$ and $D_1(\underline{\alpha}^{I},p_1,p_2)=\underline{\pi}_{11}^{(I)}\log(\frac{\underline{\pi}_{11}^{(I)}}{\underline{\pi}_{10}^{(I)}})+(1-\underline{\pi}_{11}^{(I)})\log(\frac{1-\underline{\pi}_{11}^{(I)}}{1-\underline{\pi}_{10}^{(I)}})$. We also have $g_0(\underline{\alpha}^{I},p_1,p_2)=\underline{\pi}_{10}^{(I)}(1-\underline{\pi}_{10}^{(I)})\underline{W}^2$, $g_0(\overline{\alpha}^{I},p_1,p_2)=\overline{\pi}_{10}^{(I)}(1-\overline{\pi}_{10}^{(I)})(\overline{W}^{I})^2$ and $g_1(\underline{\alpha}^{I},p_1,p_2)=\underline{\pi}_{11}^{(I)}(1-\underline{\pi}_{11}^{(I)})\underline{W}^2$, $g_1(\overline{\alpha}^{I},p_1,p_2)=\overline{\pi}_{11}^{(I)}(1-\overline{\pi}_{11}^{(I)})(\overline{W}^{I})^2$ where  $\underline{W}^{I}=\log(\frac{\underline{\pi}_{11}^{I}(1-\underline{\pi}_{10}^{I})}{\underline{\pi}_{10}^{I}(1-\underline{\pi}_{11}^{I})})$ and $\overline{W}^{I}=\log(\frac{\overline{\pi}_{11}^{I}(1-\overline{\pi}_{10}^{I})}{\overline{\pi}_{10}^{I}(1-\overline{\pi}_{11}^{I})})$. The optimal attacking strategy is stated based on \eqref{eq:gama_f_m} in the following theorem.
\begin{figure*}
\begin{subequations}\label{eq:gama_f_m}
\begin{align}
    \gamma_f^{I}&=\frac{\log(\frac{\pi_0}{\pi_1})/\sqrt{N}+\sqrt{N}(D_0(\overline{\alpha}^{I},p_1,p_2)p(u_{n}=z_{n})+D_0(\underline{\alpha}^{I},p_1,p_2)p(u_{n}\neq z_{n}))}{\sqrt{p(u_{i}\neq z_{i})g_0(\overline{\alpha}^{I},p_1,p_2)+p(u_{i}=z_{i})g_0(\underline{\alpha}^{I},p_1,p_2)}}\\
    \gamma_m^{I}&=\frac{\log(\frac{\pi_0}{\pi_1})/\sqrt{N}+\sqrt{N}(D_1(\overline{\alpha}^{I},p_1,p_2)p(u_{n}=z_{n})+D_1(\underline{\alpha}^{I},p_1,p_2)p(u_{n}\neq z_{n}))}{\sqrt{p(u_{i}\neq z_{i})g_1(\overline{\alpha}^{I},p_1,p_2)+p(u_{i}=z_{i})g_1(\underline{\alpha}^{I},p_1,p_2)}},
\end{align}
\end{subequations}
\end{figure*}
\paragraph*{Theorem 1} In the traditional audit based system, if the intelligent Byzantine attackers adopt the strategy given by $p_2=0$ when $\alpha_0\in[0,1]$, the system reduces to the one without audit bits and it can always be made blind by choosing $p_1$ such that $\alpha_0p_1=\frac{1}{2}$ if $\alpha_0\geq 0.5$.

\begin{IEEEproof}
 Please see Appendix \ref{Pf:P_e}. 
\end{IEEEproof}

\begin{figure}[htbp]
    \centering
    {\includegraphics[width=18em,height=17em]{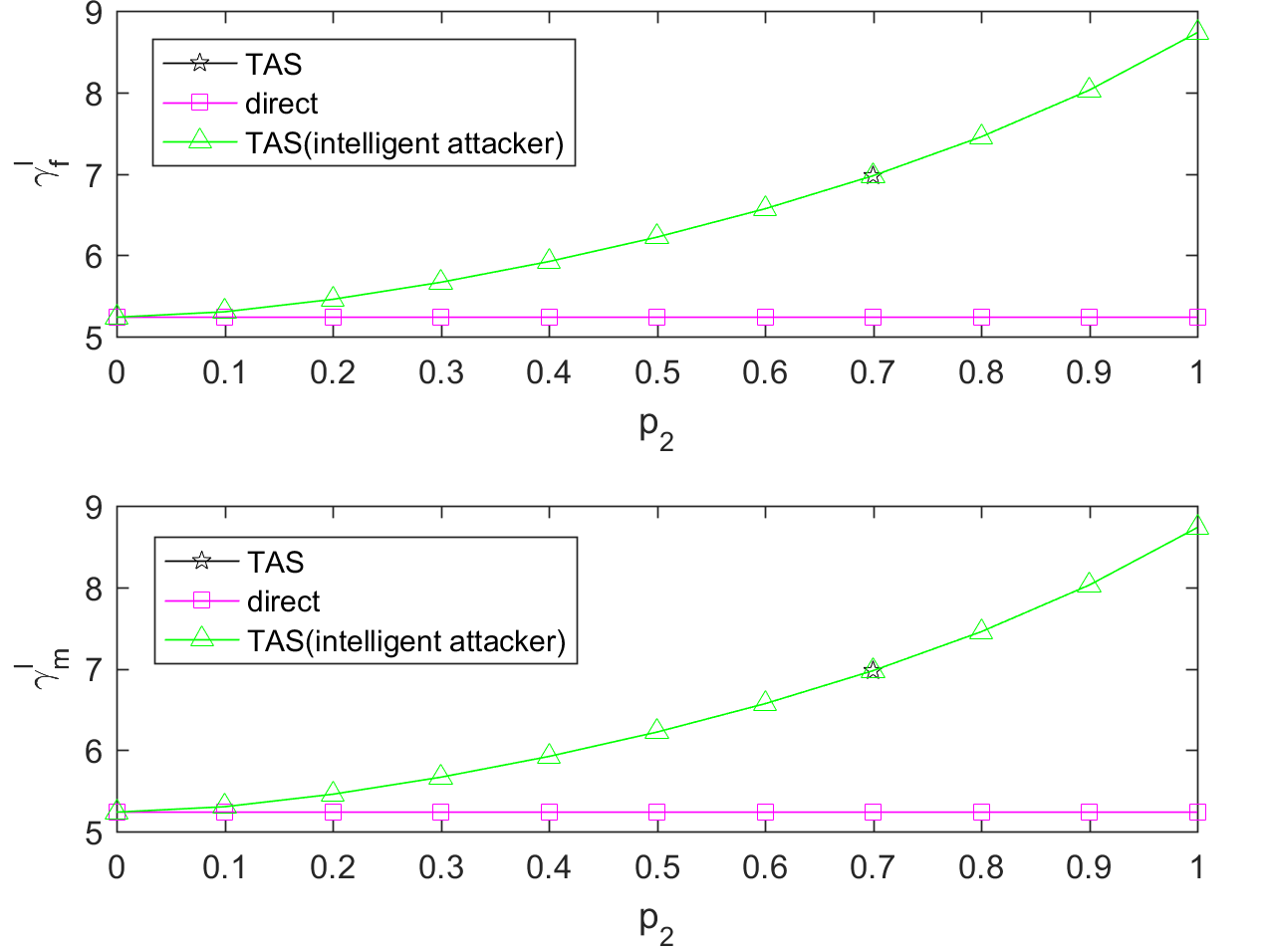}}
    \caption{$\gamma_f^{I}$ and $\gamma_m^{I}$ versus $p_2$ given $p_1=0.7$ and $\alpha_0=0.3$. Note that $p_1=p_2=0.7$ in TAS.}
    \label{gamma_I}
\end{figure}

Note that the probability of error for the system under intelligent attack is $P_e^I=\pi_0Q\left(\gamma_f^{I}\right)+\pi_1Q\left(\gamma_m^{I}\right)$. $\gamma_f^{I}$ and $\gamma_m^{I}$ are the arguments of function $Q(.)$ for the probability of false alarm and the argument of function $Q(.)$ for the probability of miss detection, respectively such that larger arguments mean better detection performance. Fig. \ref{gamma_I} shows how $\gamma_f^{I}$ and $\gamma_m^{I}$ change with $p_2$. We can observe that both $\gamma_f^{I}$ and $\gamma_m^{I}$ achieve the minimum when $p_2=0$, which means that $P_e^I$ achieves the maximum. We can also observe that arguments that attain this are equal to the ones in the system that does not use audit bits and thus $P_e^I$ reduces to the probability of error of the system that does not use audit bits. Hence, Fig. \ref{gamma_I} is in accordance with the result given in Theorem 1.

Based on the analysis above, the assumption $p_1=p_2$ given in~\cite{hashlamoun2018audit} is not the optimal choice for the attackers in practice. The attackers can launch stronger attacks when they set $p_2=0$. Under this attacking strategy, there is no improvement in the detection performance of TAS compared with the direct scheme. Thus, we conclude that the intelligent attackers can hide themselves by not flipping the decisions from their group members, i.e., $p_2=0$, according to Theorem 1 and Fig. \ref{gamma_I}. Moreover, when $p_2=0$, the detection error for TAS is the same as the one for the direct scheme. To enhance the robustness of the system, we propose a new scheme called enhanced audit bit based scheme (EAS) in next section.

\section{Enhanced Audit Bit Based Scheme}

In this section, an enhanced audit bit based scheme (EAS) is proposed to improve the robustness of the system under intelligent attacks. In TAS, the behavioral identity of each sensor is characterized by $\underline{\alpha}$ and $\overline{\alpha}$. The evaluations of the value of $\underline{\alpha}$ and $\overline{\alpha}$ only depends on its own status indicator as discussed in Section \RNum{2}. However, in the newly proposed scheme, we utilize both the status indicators of the sensors in the same group to more accurately infer the behavioral identities of sensors in the network compared with TAS.

The status indicators $\{d_i\}_{i=1}^N$ are again made by the MMSD. However, the sensors are no longer partitioned into two sets ($\underline{S}$ and $\overline{S}$). They are partitioned into four sets which are $\underline{S}\underline{S}$ $\underline{S}\overline{S}$, $\overline{S}\underline{S}$ and $\overline{S}\overline{S}$ based on both status indicators of sensor $i$ and sensor $j$ in the same group. If $d_i=d_j=1$, sensor $i$ and sensor $j$ are both placed in the set $\underline{S}\underline{S}$. If $d_i=0$ and $d_j=1$, sensor $i$ is placed in the set $\underline{S}\overline{S}$ and sensor $j$ is placed in the set $\overline{S}\underline{S}$. If $d_i=d_j=0$, sensor $i$ and sensor $j$ are both placed in the set $\overline{S}\overline{S}$. We still assume a general attacking strategy which is $p_1\neq p_2$. Then, we have the following four cases. 
\paragraph{If $i\in\underline{S}\underline{S}$} $i$ is a Byzantine node with probability

\begin{equation}\label{eq:alpha_LL}
\begin{split}
    \alpha_1=&P(i=B|i,j\in\underline{S}\underline{S})\\
    =&P(i=B,j=H|i,j\in\underline{S}\underline{S})+P(i=B,j=B|i,j\in\underline{S}\underline{S})\\
    =&\frac{P(i,j\in\underline{S}\underline{S}|i=B,j=H)P(i=B,j=H)}{P(i,j\in\underline{S}\underline{S})}\\
    &+\frac{P(i,j\in\underline{S}\underline{S}|i=B,j=B)P(i=B,j=B)}{P(i,j\in\underline{S}\underline{S})}\\
    =&\frac{\alpha_0^2f_{BB}^{(1)}+\alpha_0(1-\alpha_0)f_{BH}^{(1)}}{P(i,j\in\underline{S}\underline{S})},
\end{split}
\end{equation}
where
\begin{equation}\label{eq:LL}
\begin{split}
    P(i,j\in\underline{S}\underline{S})=&\alpha_0^2f_{BB}^{(1)}+\alpha_0(1-\alpha_0)(f_{HB}^{(1)}+f_{BH}^{(1)})\\
    &+(1-\alpha_0)^2f_{HH}^{(1)}
    \end{split}
\end{equation}
and $f_{BB}^{(1)}=[2p_1p_2(1-p_1)+(1-2p_1+2p_1^2)(1-p_2)]^2$, $f_{HB}^{(1)}=f_{BH}^{(1)}=(1-p_2)(1-2p_1+2p_1^2)$ and $f_{HH}^{(1)}=1$.
\paragraph{If $i\in\underline{S}\overline{S}$} $i$ is a Byzantine node with probability

\begin{equation}\label{eq:alpha_LU}
\begin{split}
    \alpha_2=&P(i=B|i\in\underline{S}\overline{S},j\in\overline{S}\underline{S})\\
    =&P(i=B,j=H|i\in\underline{S}\overline{S},j\in\overline{S}\underline{S})\\
    &+P(i=B,j=B|i\in\underline{S}\overline{S},j\in\overline{S}\underline{S})\\
    =&\frac{\alpha_0^2f_{BB}^{(2)}+\alpha_0(1-\alpha_0)f_{BH}^{(2)}}{P(i\in\underline{S}\overline{S},j\in\overline{S}\underline{S})},
\end{split}
\end{equation}
where
\begin{equation}
\begin{split}
    P(i\in\underline{S}\overline{S},j\in\overline{S}\underline{S})=&\alpha_0^2f_{BB}^{(2)}+\alpha_0(1-\alpha_0)(f_{HB}^{(2)}+f_{BH}^{(2)})\\
    &+(1-\alpha_0)^2f_{HH}^{(2)}
    \end{split}
\end{equation}
and $f_{BB}^{(2)}=[2p_1p_2(1-p_1)+(1-2p_1+2p_1^2)(1-p_2)][1-2p_1p_2(1-p_1)-(1-2p_1+2p_1^2)(1-p_2)]$, $f_{HB}^{(2)}=p_2(1-2p_1+2p_1^2)$,$f_{BH}^{(2)}=2p_1(1-p_2)(1-p_1)$ and $f_{HH}^{(2)}=0$.
\paragraph{If $i\in\overline{S}\underline{S}$} $i$ is a Byzantine node with probability
\begin{equation}\label{eq:alpha_UL}
\begin{split}
    \alpha_3=&P(i=B|i\in\overline{S}\underline{S},j\in\underline{S}\overline{S})\\
    =&P(i=B,j=H|i\in\overline{S}\underline{S},j\in\underline{S}\overline{S})\\
    &+P(i=B,j=B|i\in\overline{S}\underline{S},j\in\underline{S}\overline{S})\\
    =&\frac{\alpha_0^2f_{BB}^{(3)}+\alpha_0(1-\alpha_0)f_{BH}^{(3)}}{P(i\in\overline{S}\underline{S},j\in\underline{S}\overline{S})},
\end{split}
\end{equation}
where
\begin{equation}
\begin{split}
    P(i\in\overline{S}\underline{S},j\in\underline{S}\overline{S})=&\alpha_0^2f_{BB}^{(3)}+\alpha_0(1-\alpha_0)(f_{HB}^{(3)}+f_{BH}^{(3)})\\
    &+(1-\alpha_0)^2f_{HH}^{(3)}
    \end{split}
\end{equation}
and $f_{BB}^{(3)}=[2p_1p_2(1-p_1)+(1-2p_1+2p_1^2)(1-p_2)][1-2p_1p_2(1-p_1)-(1-2p_1+2p_1^2)(1-p_2)]$, $f_{HB}^{(3)}=2p_1(1-p_2)(1-p_1)$,$f_{BH}^{(3)}=p_2(1-2p_1+2p_1^2)$ and $f_{HH}^{(3)}=0$.

\paragraph{If $i\in\overline{S}\overline{S}$} $i$ is a Byzantine node with probability

\begin{equation}\label{eq:alpha_UU}
\begin{split}
    \alpha_4=&P(i=B|i,j\in\overline{S}\overline{S})\\
    =&P(i=B,j=H|i,j\in\overline{S}\overline{S})\\
    &+P(i=B,j=B|i,j\in\overline{S}\overline{S})\\
    =&\frac{\alpha_0^2f_{BB}^{(4)}+\alpha_0(1-\alpha_0)f_{BH}^{(4)}}{P(i,j\in\overline{S}\overline{S})},
\end{split}
\end{equation}
where
\begin{equation}
\begin{split}
    P(i,j\in\overline{S}\overline{S})=&\alpha_0^2f_{BB}^{(4)}+\alpha_0(1-\alpha_0)(f_{HB}^{(4)}+f_{BH}^{(4)})\\
    &+(1-\alpha_0)^2f_{HH}^{(4)}
    \end{split}
\end{equation}
and $f_{BB}^{(4)}=[2p_1(1-p_2)(1-p_1)+p_2p_1^2]^2$, $f_{HB}^{(4)}=f_{BH}=2p_1p_2(1-p_1)$ and $f_{HH}^{(4)}=0$.

The next lemma shows that our proposed EAS performs a more accurate evaluation of the behavioral identity of each sensor compared with TAS.
\paragraph*{Lemma 2}
The probability of sensor $i$ being a Byzantine node when $i\in\underline{S}$ in TAS is equal to the weighted average of the probabilities of sensor $i$ being a Byzantine node when $i,j\in\underline{S}\underline{S}$ and $i\in\underline{S}\overline{S},j\in\overline{S}\underline{S}$, respectively. That is
\begin{equation}\label{eq:proof_aver}
\begin{split}
    &P(i=B|i\in\underline{S})\\
    &=\alpha_1P(d_j=1|d_i=1)+\alpha_2P(d_j=0|d_i=1)
    \end{split}
\end{equation}
A similar result can be obtained for sensor $i\in\overline{S}$.

\begin{IEEEproof}
 The right hand side (RHS) of \eqref{eq:proof_aver} is the same as $P(i=B|d_i=1,d_j=1)P(d_j=1|d_i=1)+P(i=B|d_i=1,d_j=0)P(d_j=0|d_i=1)$.
According to the Bayes' rule, we have
\begin{equation}\label{eq:prove_aver}
\begin{split}
    &\sum_{x=0,1}P(i=B|d_i=1,d_j=x)P(d_j=x|d_i=1)\\
    &=\sum_{x=0,1}P(i=B|d_i=1,d_j=x)\frac{P(d_i=1,d_j=x)}{P(d_i=1)}\\
    &=\sum_{x=0,1}\frac{\alpha_0^2f_{BB}^{(x)}+\alpha_0(1-\alpha_0)f_{BH}^{(x)}}{P(d_i=1,d_j=x)}\frac{P(d_i=1,d_j=x)}{P(d_i=1)}\\
    &=\sum_{x=0,1}\frac{\alpha_0^2f_{BB}^{(x)}+\alpha_0(1-\alpha_0)f_{BH}^{(x)}}{P(d_i=1)}\\
    &=P(i=B|i\in\underline{S})
\end{split}
\end{equation}

We can also show that $i\in\overline{S}$ is the weighted average of the probabilities of sensor $i$ being a Byzantine node when $i,j\in\overline{S}\overline{S}$ and $i\in\overline{S}\underline{S},j\in\underline{S}\overline{S}$ by following a similar procedure and, therefore, the details of its proof are omitted here.
\end{IEEEproof}

\begin{figure}
    \centering
    \includegraphics[width=18em,height=11em]{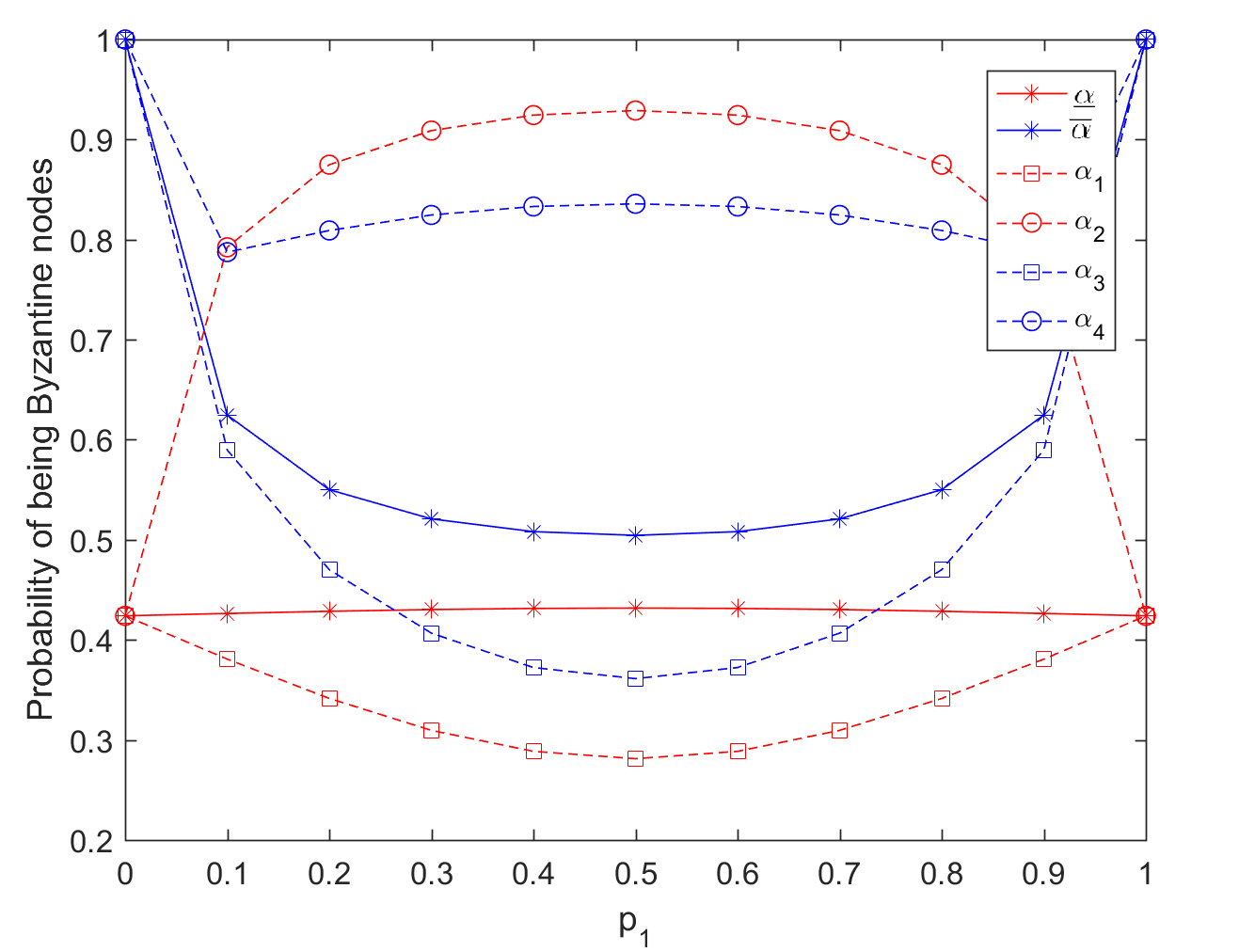}
    \caption{The probability of being Byzantine nodes for sensors in sets $\underline{S}$, $\overline{S}$, $\underline{S}\underline{S}$, $\underline{S}\overline{S}$, $\overline{S}\underline{S}$ and $\overline{S}\overline{S}$ when $p_2=0.1$.}
    \label{fig:aver}
\end{figure}

Fig. \ref{fig:aver} corroborates the results in Lemma 2. Note that each sensor placed in $\underline{S}$ (or $\overline{S}$) is a Byzantine node with probability of $\underline{\alpha}$ (or $\overline{\alpha}$) for TAS. We can observe that the value of $\underline{\alpha}$ (or $\overline{\alpha}$) is in the middle of the values of $\alpha_1$ and $\alpha_2$ (or $\alpha_3$ and $\alpha_4$) for the proposed scheme. It shows that taking both the status indicators from the same group into consideration can give us more information about the the behavioral identities of the sensors in the network. Hence, our proposed EAS outperforms TAS that only utilizes the averaged probabilities ($\underline{\alpha}$ or $\overline{\alpha}$) to assess the behavioral identity for each sensor. Thus, the pmf of local decision $u_i$ for our proposed EAS is expressed as
\begin{equation}\label{eq:P_ui_enhan}
    P(u_i|\mathcal{H}_q)=
\begin{cases}
\pi_{1q,1}^{u_i}(1-\pi_{1q,1})^{1-{u_i}}& \text{for $i\in\underline{S}\underline{S}$}\\
\pi_{1q,2}^{u_i}(1-\pi_{1q,2})^{1-{u_i}}& \text{for $i\in\underline{S}\overline{S}$}\\
\pi_{1q,3}^{u_i}(1-\pi_{1q,3})^{1-{u_i}}& \text{for $i\in\overline{S}\underline{S}$}\\
\pi_{1q,4}^{u_i}(1-\pi_{1q,4})^{1-{u_i}}& \text{for $i\in\overline{S}\overline{S}$}
\end{cases}
\end{equation}
for $q=0,1$, where
\begin{subequations}\label{eq:pi_11_10_under_en}
\begin{align}
    &\pi_{11,e}=1-\pi_{10,e}=P_d(1-\alpha_ep_1)+\alpha_ep_1(1-P_d)\\
    &\pi_{10,e}=1-\pi_{00,e}=P_f(1-\alpha_ep_1)+\alpha_ep_1(1-P_f)
\end{align}
\end{subequations}
for $e=1,2,3,4$. $\pi_{11,e}$ and $\pi_{10,e}$ are the probabilities of sending the local decision $u_i=1$ given hypothesis $\mathcal{H}_1$ and given hypothesis $\mathcal{H}_0$, respectively, for $e=1,2,3,4$ which are corresponding to the sensors being in $\underline{S}\underline{S}$ $\underline{S}\overline{S}$, $\overline{S}\underline{S}$ and $\overline{S}\overline{S}$. The new optimal decision rule is provided in Theorem 2.

\paragraph*{Theorem 2} The new decision rule for the proposed EAS, given the Byzantine flipping probabilities $p_1$, $p_2$ and $\alpha_0$ fraction of Byzantine nodes, is expressed as

\begin{equation}\label{eq:LLR_enhan}
\begin{split}
    \sum_{e=1}^4 W_{e}U_e\gtrless \eta^{(En)},
\end{split}
\end{equation}
where $U_1=\sum_{i\in\underline{S}\underline{S}}u_i$, $U_2=\sum_{i\in\underline{S}\overline{S}}u_i$, $U_3=\sum_{i\in\overline{S}\underline{S}}u_i$, $U_4=\sum_{i\in\overline{S}\overline{S}}u_i$ and $W_e=\log(\frac{\pi_{11,e}(1-\pi_{10,e})}{\pi_{10,e}(1-\pi_{11,e})})$ for $e=1,2,3,4$. $\eta^{(En)}$ is the threshold used by the FC for EAS, where $\eta^{(En)}=\log(\frac{\pi_0}{\pi_1})+\sum_{e=1}^4N_e\log(\frac{1-\pi_{10,e}}{1-\pi_{11,e}})$. $N_1$, $N_2$, $N_3$ and $N_4$ are the cardinalities of sets $\underline{S}\underline{S}$, $\underline{S}\overline{S}$, $\overline{S}\underline{S}$ and $\overline{S}\overline{S}$, respectively, where $N_1=|\underline{S}\underline{S}|$ ,$N_2=|\underline{S}\overline{S}|$, $N_3=|\overline{S}\underline{S}|$ and $N_4=|\overline{S}\overline{S}|$.

\begin{IEEEproof}
 We know that the local decisions are independent given the hypothesis $\mathcal{H}_0$ or $\mathcal{H}_1$ and the information about the sets where all the sensors are placed in. Hence, the optimal decision rule, which is given in \eqref{eq:proof_LLR_enhan}, can be further simplified. Substituting \eqref{eq:P_ui_enhan} in \eqref{eq:proof_LLR_enhan}, and taking the logarithm on both sides, we obtain the fusion rule in the theorem.
\begin{figure*}[ht]
\begin{equation}\label{eq:proof_LLR_enhan}
    \prod_{i\in\underline{S}\underline{S}}\frac{P(u_i|\mathcal{H}_1)}{P(u_i|\mathcal{H}_0)}\prod_{i\in\underline{S}\overline{S}}\frac{P(u_i|\mathcal{H}_1)}{P(u_i|\mathcal{H}_0)}\prod_{i\in\overline{S}\underline{S}}\frac{P(u_i|\mathcal{H}_1)}{P(u_i|\mathcal{H}_0)}\prod_{i\in\overline{S}\overline{S}}\frac{P(u_i|\mathcal{H}_1)}{P(u_i|\mathcal{H}_0)}\gtrless \frac{\pi_0}{\pi_1}
\end{equation}
\end{figure*}
\end{IEEEproof}

\begin{figure}[htb]
  \centering
  \subfigure[$\gamma_f$ as a function of flipping probability $p_1$ given $p_2=0$ and $p_2=0.2$.]{
    \includegraphics[width=18em,height=11em]{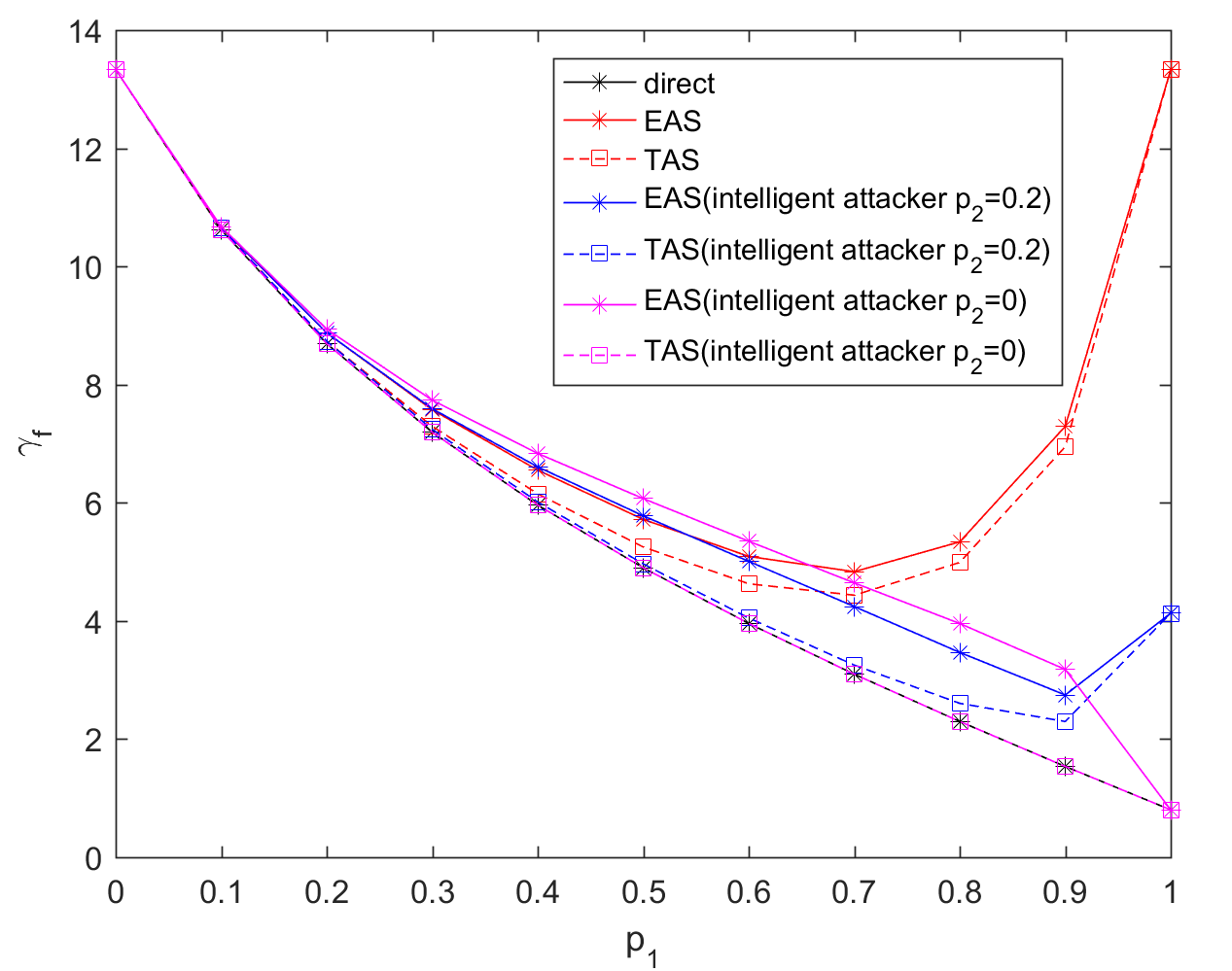}}
      \subfigure[$\gamma_m$ as a function of flipping probability $p_1$ given $p_2=0$ and $p_2=0.2$.]{
    \includegraphics[width=18em,height=11em]{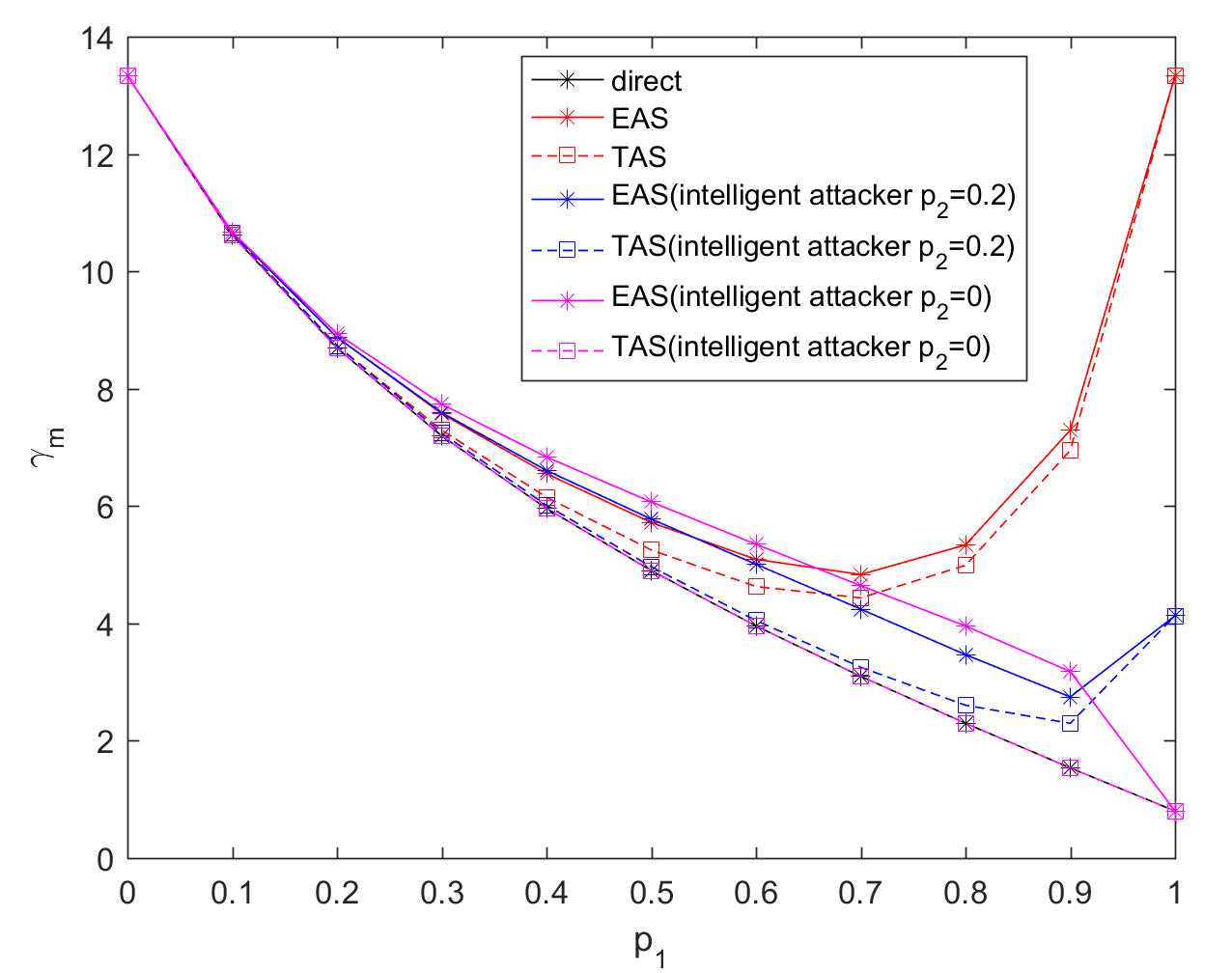}}
  \caption{The probability of error is characterized by the argument of function $Q(.)$ for the probability of false alarm shown in (a) and the argument of function $Q(.)$ for the probability of miss detection shown in (b). Smaller values of the argument result in higher probabilities of error.}
  \label{fig:enhanced}
\end{figure}

Note that $U_e$ is binomial distributed random variables with parameters $(N,\pi_{11,e})$ under $\mathcal{H}_1$, and with parameters $(N,\pi_{10,e})$ under $\mathcal{H}_0$ for $e=1,2,3,4$. When $N$ is large, $N_1$, $N_2$, $N_3$ and $N_4$ can be approximated by their expected value $NP(i\in\underline{S}\underline{S})$, $NP(i\in\underline{S}\overline{S})$, $NP(i\in\overline{S}\underline{S})$ and $NP(i\in\overline{S}\overline{S})$, respectively. For any sensor $i\in \{1,2,\dots,N\}$, the probability of being placed in $\underline{S}\underline{S}$, $\underline{S}\overline{S}$, $\overline{S}\underline{S}$ and $\overline{S}\overline{S}$ are $P(i\in\underline{S}\underline{S})=P(d_i=d_j=1)$, $P(i\in\underline{S}\overline{S})=P(i\in\overline{S}\underline{S})=P(d_i=1,d_j=0)=P(d_i=0,d_j=1)$ and $P(i\in\overline{S}\overline{S})=P(d_i=d_j=0)$, respectively. The threshold used by the FC becomes $\eta^{(En)}=\log(\frac{\pi_0}{\pi_1})+NP(i\in\underline{S}\underline{S})\log(\frac{1-\pi_{10,1}}{1-\pi_{11,1}})+NP(i\in\underline{S}\overline{S})\log(\frac{1-\pi_{10,2}}{1-\pi_{11,2}})+NP(i\in\overline{S}\underline{S})\log(\frac{1-\pi_{10,3}}{1-\pi_{11,3}})+NP(i\in\overline{S}\overline{S})\log(\frac{1-\pi_{10,4}}{1-\pi_{11,4}})$. Thus, the PDF of the global static $U=\sum_{e=1}^4 W_{e}U_e$ can be approximated by the Gaussian distribution with parameters given as follows.
\begin{small}
\begin{subequations}\label{eq:mu_sigma_en}
    \begin{align}
        \mu_0^{(En)}=&E[U|\mathcal{H}_0]\notag\\
        =&N(P(i\in\underline{S}\underline{S})\pi_{10,1}W_1+P(i\in\underline{S}\overline{S})\pi_{10,2}W_2\notag\\
        &+ P(i\in\overline{S}\underline{S})\pi_{10,3}W_3+P(i,j\in\overline{S}\overline{S})\pi_{10,4}W_4)\\
        \mu_1^{(En)}=&E[U|\mathcal{H}_1]\notag\\
        &=N(P(i\in\underline{S}\underline{S})\pi_{11,1}W_1+P(i\in\underline{S}\overline{S})\pi_{11,2}W_2\notag\\
        &+ P(i\in\overline{S}\underline{S})\pi_{11,3}W_3+P(i\in\overline{S}\overline{S})\pi_{11,4}W_4)\\
        (\sigma_0^{(En)})^2=&Var[U|\mathcal{H}_0]\notag\\
        =&N(P(i\in\underline{S}\underline{S})\pi_{10,1}(1-\pi_{10,1})W_1^2\notag\\
        &+P(i\in\underline{S}\overline{S})\pi_{10,2}(1-\pi_{10,2})W_2^2\notag\\
        &+ P(i\in\overline{S}\underline{S})\pi_{10,3}(1-\pi_{10,3})W_3^2\notag\\
        &+P(i\in\overline{S}\overline{S})\pi_{10,4}(1-\pi_{10,4})W_4^2)\\
        (\sigma_1^{(En)})^2=&Var[U|\mathcal{H}_1]\notag\\
        &=N(P(i\in\underline{S}\underline{S})\pi_{11,1}(1-\pi_{11,1})W_1^2\notag\\
        &+P(i\in\underline{S}\overline{S})\pi_{11,2}(1-\pi_{11,2})W_2^2\notag\\
        &+ P(i\in\overline{S}\underline{S})\pi_{11,3}(1-\pi_{11,3})W_3^2\notag\\
        &+P(i\in\overline{S}\overline{S})\pi_{11,4}(1-\pi_{11,4})W_4^2)
    \end{align}
\end{subequations}
\end{small}
The detection performance, characterized by the probability of error of the system, is given as
\begin{equation}\label{eq:pe_en}
    P_e^{(En)}=\pi_0Q\left(\gamma_f^{(En)}\right)+\pi_1Q\left(\gamma_m^{(En)}\right),
\end{equation}

where $\gamma_f^{(En)}=\frac{\eta^{(En)}-\mu_0^{(En)}}{\sigma_0^{(En)}}$ and $\gamma_m^{(En)}=\frac{\mu_1^{(En)}-\eta^{(En)}}{\sigma_1^{(En)}}$. Fig. \ref{fig:enhanced} shows that the detection performance of the proposed scheme in terms of $\gamma_f^{(En)}$ and $\gamma_m^{(En)}$ is better than the detection performance of the traditional one, TAS, under both intelligent attacks and non-intelligent attacks. We can observe that the detection performance of TAS is the same as the direct scheme when the system is under intelligent attacks ($p_2=0$). This is in accordance with the results shown in Theorem 1. However, the proposed EAS prevents it from happening.
As shown in Fig. \ref{fig:enhanced}, the worst case from the perspective of the FC is that the intelligent attackers take the attacking strategy of $p_1=1$ and $p_2=0$. In this case, the proposed EAS has the same detection performance as the direct scheme. In the next section, another new scheme is proposed which achieves better detection performance and higher robustness compared with EAS.

\section{Proposed Optimal Bayesian Fusion Rule}

In this section, we propose a new framework and a new fusion rule for the audit bit based system. In this framework, we focus on the practical scenario in which the Byzantine nodes are in a minority due to the limited attacking resources, i.e., $\alpha_0\leq1/2$. We will first start with a network with one cluster, then we will move on to a wide-area network with multiple clusters.
\subsection{A single-cluster network}
As before, the sensors are partitioned into sets $\underline{S}\underline{S}$, $\underline{S}\overline{S}$, $\overline{S}\underline{S}$ and $\overline{S}\overline{S}$ by the MMSD based on both status indicators of sensor $i$  and sensor $j$ in the same group. Moreover, the local decisions $(u_{i},u_{j})$ sent from the same group are also compared to give us additional information about the behavioral identity of sensors in the networks. Each sensor again transmits its decision to the MMSD via two paths, namely the direct path and indirect path to the FC. After collecting all the local decisions, the MMSD places the sensors into sets $\underline{S}\underline{S}$, $\underline{S}\overline{S}$, $\overline{S}\underline{S}$ and $\overline{S}\overline{S}$. These steps are the same as the ones in EAS. However, the MMSD also considers the MMS of the decisions $u_{i}$ and $u_{j}$ from the same group: if the sensor decisions for sensors $i$ and $j$ are the same, i. e., $u_{i}=u_{j}$, they are placed in the Set $\underline{\mathcal{M}}$ and the others are placed in the Set $\overline{\mathcal{M}}$. The MMSD only transmits the local decisions of the sensors with the sensor index $i$ given by  $\{i:\left(\underline{S}\underline{S}\bigcap\underline{\mathcal{M}}\right)\bigcup\underline{S}\overline{S}\bigcup\overline{S}\underline{S}\}$ to the decision making module to make the final decision. In other words, the local decisions from the sensors in Set 
$\underline{S}\underline{S}\bigcap\overline{\mathcal{M}}$ or Set $\overline{S}\overline{S}$ are not used to make the final decision which correspond to the two conditions stated as below.
\paragraph*{Condition 1} The sensor $i$ and its group member $j$ are both in the set $\overline{S}\overline{S}$.
\paragraph*{Condition 2} The sensor $i$ and its group member $j$ are both in the set $\underline{S}\underline{S}$ and $u_{i}\neq u_{j}$.

In the next lemma, we show the reasons why not using the decisions of sensors that satisfy one of the above two conditions improves the detection performance of the system.
\paragraph*{Lemma 3}
\begin{enumerate}
	\item  When the sensor pair $(i,j)$ satisfies Condition
    1, i. e., sensors $i$ and$j$ belong to $\overline{S}\overline{S}$, removing this sensor pair results in the removal of two Byzantine nodes when $p_2=0$.
	\item  When we remove the sensor pairs that
    satisfy Condition 2, the ability of removing the Byzantine nodes for the proposed RAS increases with the increase of $p_1$ given specific $p_2$ and $\alpha_0$. 
\end{enumerate}
\begin{IEEEproof}
 \begin{enumerate}
	\item Let $E$ be the event that at least one node in sensor pair $(i,j)$ is a Byzantine node. When $i,j\in \overline{S}\overline{S}$, it is obvious that $P(E|i,j\in \overline{S}\overline{S})=1$. Thus, we can obtain $P(i,j\notin \overline{S}\overline{S}|\overline{E})=1$ due to the fact that the contrapositive of the conditional statement is also true. So we can conclude that there is at least one Byzantine node in the sensor pair. Moreover, it is easy to conclude that all the sensors are Byzantine nodes in the Set $\overline{S}\overline{S}$ when the attackers take the strategy of $p_2=0$  according to $\eqref{eq:alpha_UU}$. Thus, removing the decisions of sensors in this set can remove at least one Byzantine node in each pair, and it can even remove two Byzantine nodes in each pair when the attackers employ the strategy of $p_2=0$.
	\item To evaluate the impact of removing the unequal local decisions of sensor pairs on the performance of removing Byzantine nodes, we utilize the ratio $F=\frac{P(E,u_i=u_j|i,j\in\underline{S}\underline{S})}{P(E|i,j\in\underline{S}\underline{S})}$ to characterize that performance. The numerator of ratio $F$ is the probability of the joint event that there exists at least one Byzantine node and the event $u_i=u_j$ given $i,j\in\underline{S}\underline{S}$. The denominator is the probability of at least one Byzantine node given $i,j\in\underline{S}\underline{S}$. The ratio $F=P(u_i=u_j|i,j\in\underline{S}\underline{S},E)$ gives the probability of $u_i=u_j$ given event $E$ and $i,j\in\underline{S}\underline{S}$.  We have
\begin{subequations}
\begin{align}
    &P(E,u_i=u_j|i,j\in\underline{S}\underline{S})\notag\\
    &=P(E|u_i=u_j,i,j\in\underline{S}\underline{S})P(u_i=u_j|i,j\in\underline{S}\underline{S})\\
    &=(1-P(i=H,j=H|i,j\in\underline{S}\underline{S},u_i=u_j))\notag\\
    &\quad \times P(u_i=u_j,|i,j\in\underline{S}\underline{S})\\
    &=P(u_i=u_j|i,j\in\underline{S}\underline{S})-P(u_i=u_j|i,j\in\underline{S}\underline{S},\notag\\
    &\quad i=H,j=H)P(i=H,j=H|i,j\in\underline{S}\underline{S})\\
    &=P(u_i=u_j|i,j\in\underline{S}\underline{S})-\frac{(1-\alpha_0)^2}{P(i,j\in\underline{S}\underline{S})}\notag\\
    &\quad \times P(u_i=u_j|i,j\in\underline{S}\underline{S},i=H,j=H)
\end{align}
\end{subequations}
and
\begin{subequations}
   \begin{align}
    &P(E|i,j\in\underline{S}\underline{S})\notag\\
    &=1-P(i=H,j=H|i,j\in\underline{S}\underline{S})\\
    &=1-\frac{(1-\alpha_0)^2}{P(i,j\in\underline{S}\underline{S})},
\end{align} 
\end{subequations}
where $P(u_i=u_j|i,j\in\underline{S}\underline{S})=P(u_{i}=u_{j}|i,j\in\underline{S}\underline{S},\mathcal{H}_0)P(\mathcal{H}_0)+P(u_{i}=u_{j}|i,j\in\underline{S}\underline{S},\mathcal{H}_1)P(\mathcal{H}_1)=\pi_1[\pi_{11,1}^2+(1-\pi_{11,1})^2]+\pi_0[\pi_{10,1}^2+(1-\pi_{10,1})^2]$ and $P(u_i=u_j|i,j\in\underline{S}\underline{S},i=H,j=H)=[P_d^2+(1-P_d)^2]\pi_1+[P_f^2+(1-P_f)^2]\pi_0$.
\end{enumerate}
\end{IEEEproof} 

\begin{figure}[htbp]
    \centering
    {\includegraphics[width=18em,height=11em]{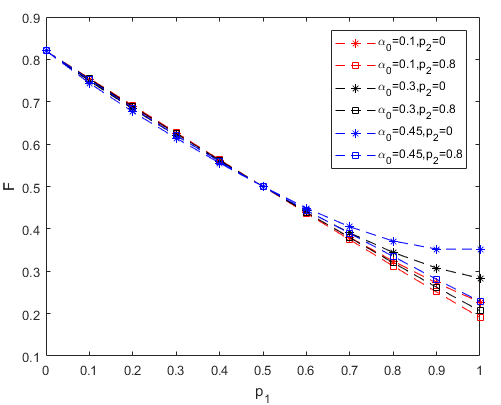}}
    \caption{$F$ versus $p_1$ given $p_2=0.1$ for different $\alpha_0$ and $N=100$.}
    \label{fig:ratio2}
\end{figure}

The relationship among $p_1$, $p_2$, $\alpha_0$ and $F$ is shown in Fig. \ref{fig:ratio2}. We can observe that the value of $F$ has a significant decrease when $p_1$ is large. It can also be observed that the value of $F$ decreases with the increase of $\alpha_0$ given $p_1\geq0.5$ and a specific $p_2$. If the value of $F$ is small, it means a lower probability of existence of Byzantine nodes in the sensor pair given $i,j\in\underline{\mathcal{M}}\bigcap\underline{S}\underline{S}$. Obviously, by removing the sensor pairs which satisfy Condition 2, the ability of removing the Byzantine nodes increases with the increase of $p_1$ for a given $p_2$.

According to Theorem 1, the attackers' optimal attacking strategy in TAS is to choose $p_2=0$. In the scenario where $p_2$ is very small (close to 0), however, Fig. \ref{gamma_I} has shown that the detection performance of TAS significantly degrades for a large value of $p_1$.
The proposed scheme in this section achieves better detection performance compared with TAS when the attackers adopt the strategy of $p_2=0$ with $\forall p_1\in[0,1]$. It is because when $p_2$ is small, the Byzantine nodes have high probabilities of being placed in the set $\underline{S}\underline{S}$ in our proposed scheme. If the attacker chooses $p_1$ to be large, there is a high probability that the group containing a Byzantine node satisfies Condition 2. Hence, the decision of the Byzantine node is likely to be blocked by the MMSD and not transmitted to the FC. As a result, our scheme prevents the attacker from designing $p_1$ to be very large and $p_2$ to be very small. On the other hand, when $p_1$ is not so large, each Byzantine node has a relatively higher probability, i.e., $1-p_1$, to act honestly. Through such a trade off, the detection accuracy of the proposed scheme outperforms TAS under intelligent attacks. 

Based on the analysis above, we can show that the proposed scheme can effectively remove the decisions coming from Byzantine nodes. Hence, in the proposed RAS, we  have the following relations for sensor $i$.
\begin{subequations}\label{eq:pi_11}
\begin{align}
    P(u_{i}=1|i\in\underline{S}\overline{S},\mathcal{H}_q)&=\pi_{1q,2}\\
    P(u_{i}=1|i\in\overline{S}\underline{S},\mathcal{H}_q)&=\pi_{1q,3}
\end{align}
\end{subequations}
where $q=0,1$. Although $u_{i}$ and $u_{j}$ are dependent given $i,j\in\underline{S}\underline{S}\bigcap\underline{\mathcal{M}}$, they are independent given $i,j\in\underline{S}\underline{S}$. Hence, we have
\begin{equation}\label{eq:pi_11_re}
\begin{split}
    &\quad P(u_{i}=1,u_{j}=1|i,j\in\underline{S}\underline{S}\bigcap\underline{\mathcal{M}},\mathcal{H}_q)\\
    &=\frac{P(u_{i}=1|i,j\in\underline{S}\underline{S},\mathcal{H}_q)P(u_{j}=1|i,j\in\underline{S}\underline{S},\mathcal{H}_q)}{P(u_{i}=u_{j}|i,j\in\underline{S}\underline{S},\mathcal{H}_q)}\\
    &=\frac{\underline{\pi}_{1q}^2}{\underline{\pi}_{1q}^2+(1-\underline{\pi}_{1q})^2}=\pi_{1q,5}
\end{split}
\end{equation}
for $q=0,1$. To simplify the analysis, we consider the group votes instead of the individual votes for the sensors in set $\underline{S}\underline{S}\bigcap\underline{\mathcal{M}}$. Let $z_g$ denote the group vote for group $g\in\underline{T}$, where $\underline{T}$ is the set of group whose sensors are in set $\underline{S}\underline{S}\bigcap\underline{\mathcal{M}}$. Due to the fact that the sensors in the same group in set $\underline{S}\underline{S}\bigcap\underline{\mathcal{M}}$ has the same decisions, we have $z_g=\{0,2\}$. Hence, we obtain the following pdfs 
\begin{equation}\label{eq:u_i}
    f(u_i|\mathcal{H}_q)=
\begin{cases}
\pi_{1q,2}^{u_i}(1-\pi_{1q,2})^{1-{u_i}}& \text{for $i\in\underline{S}\overline{S}$}\\
\pi_{1q,3}^{u_i}(1-\pi_{1q,3})^{1-{u_i}}& \text{for $i\in\overline{S}\underline{S}$}
\end{cases}
\end{equation}
for sensor $i\in\underline{S}\overline{S}\bigcup\underline{S}\overline{S}$, and
\begin{equation}\label{eq:u_i2}
    f(z_g|\mathcal{H}_q)=
\pi_{1q,5}^{z_g/2}(1-\pi_{1q,5})^{1-{z_g/2}}
\end{equation}
for group $g\in\underline{T}$, where $q=0,1$.
Thus, the proposed new decision rule is shown in Theorem 2.

\paragraph*{Theorem 2} The new optimal decision rule, given the Byzantine flipping probabilities $p_1$, $p_2$ and $\alpha_0$ fraction of Byzantine nodes, is expressed as

\begin{equation}\label{eq:LLR_simple2}
\begin{split}
    W_{5}\sum_{g\in\underline{T}}\frac{z_g}{2}+W_{2}\sum_{i\in\underline{S}\overline{S}}u_i+W_{3}\sum_{i\in\overline{S}\underline{S}}u_i\gtrless \eta^{(RA)},
\end{split}
\end{equation}
where $W_{2}=\log(\frac{\pi_{11,2}(1-\pi_{10,2})}{\pi_{10,2}(1-\pi_{11,2})})$,  $W_{3}=\log(\frac{\pi_{11,3}(1-\pi_{10,3})}{\pi_{10,3}(1-\pi_{11,3})})$, $\eta^{(RA)}=\log(\frac{\pi_0}{\pi_1})+N_{re}^{LL}\log(\frac{1-\pi_{10,5}}{1-\pi_{11,5}})+N_{re}^{L}\log(\frac{1-\pi_{10,2}}{1-\pi_{11,2}})+N_{re}^{U}\log(\frac{1-\pi_{10,3}}{1-\pi_{11,3}})$. $N^L_{re}$, $N^U_{re}$ and $N_{re}^{LL}$ are the cardinalities of sets $\underline{S}\overline{S}$, $\overline{S}\underline{S}$ and $\underline{T}$, respectively, where $N^L_{re}=|\underline{S}\overline{S}|$,$N^U_{re}=|\overline{S}\underline{S}|$, and $N_{re}^{LL}=|\underline{T}|$. $W_{5}$ denotes the rearranged weight for group decisions in set $\underline{T}$ which is given as
\begin{equation}
    W_{5}=\frac{\pi_{11,5}(1-\pi_{10,5})}{\pi_{10,5}(1-\pi_{11,5})}.
\end{equation}

\begin{IEEEproof}
We know that all groups of sensors whose decisions are sent to the FC are elements of one of the three sets $\underline{S}\overline{S}$, $\overline{S}\underline{S}$ and $\underline{S}\underline{S}\bigcap\underline{\mathcal{M}}$. Thus, the optimal decision rule is given as \eqref{eq:proof_LLR} due to the fact that the sensors in sets $\underline{S}\overline{S}$ or $\overline{S}\underline{S}$ independently send their local decisions to the FC given the hypothesis $\mathcal{H}_0$ or $\mathcal{H}_1$.
Even though the decisions coming from the sensors in the same group in set $\underline{S}\underline{S}\bigcap\underline{\mathcal{M}}$ are dependent, the group votes are independent of each other. Hence, the optimal decision rule can be reformulated as \eqref{eq:proof_LLR2}.  Substituting \eqref{eq:pi_11}, \eqref{eq:pi_11_re}, \eqref{eq:u_i}, \eqref{eq:u_i2} in \eqref{eq:proof_LLR2}, and taking the logarithm on both sides, we can get the fusion rule stated in the theorem.
\begin{figure*}[ht]
\begin{equation}\label{eq:proof_LLR}
    \prod_{i,j\in\underline{S}\underline{S}\bigcap\underline{\mathcal{M}}}\frac{P(u_i,u_j|\mathcal{H}_1)}{P(u_i,u_j|\mathcal{H}_0)}\prod_{i\in\underline{S}\overline{S}}\frac{P(u_i|\mathcal{H}_1)}{P(u_i|\mathcal{H}_0)}\prod_{i\in\overline{S}\underline{S}}\frac{P(u_i|\mathcal{H}_1)}{P(u_i|\mathcal{H}_0)}\gtrless \frac{\pi_0}{\pi_1}
\end{equation}
\end{figure*}
\begin{equation}\label{eq:proof_LLR2}
    \prod_{g\in\underline{T}}\frac{P(z_g|\mathcal{H}_1)}{P(z_g|\mathcal{H}_0)}\prod_{i\in\underline{S}\overline{S}}\frac{P(u_i|\mathcal{H}_1)}{P(u_i|\mathcal{H}_0)}\prod_{i\in\overline{S}\underline{S}}\frac{P(u_i|\mathcal{H}_1)}{P(u_i|\mathcal{H}_0)}\gtrless \frac{\pi_0}{\pi_1}
\end{equation}
\end{IEEEproof}

\begin{figure}[htb]
  \centering
  \subfigure[$\gamma_f$ as a function of flipping probability $p_1$ given $p_2=0$ and $p_2=0.2$ when $\alpha_0=0.45$.]{
    \includegraphics[width=18em,height=11em]{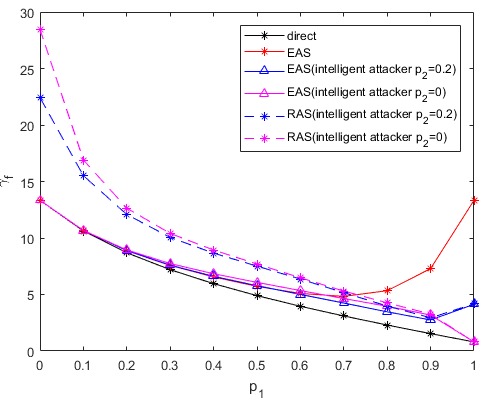}}
      \subfigure[$\gamma_f$ as a function of flipping probability $p_1$ given $p_2=0$ and $p_2=0.2$ when $\alpha_0=0.15$.]{
    \includegraphics[width=18em,height=11em]{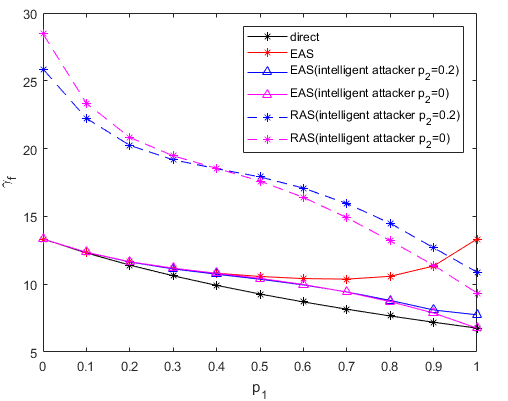}}
  \caption{The argument for the probability of false alarm function for different values of $\alpha_0$.}
  \label{fig:gamma_alp_0.15_0}
\end{figure}

Let $U$ denote the left-hand side of the optimal decision rule in \eqref{eq:LLR_simple2} which is given as
\begin{equation}\label{eq:LLR_simple_lhs}
    U=W_{5}U_{5}+W_{2}U_{2}+W_{3}U_{3},
\end{equation}
where $U_{5}=\sum_{g\in\underline{T}}z_g/2$, $U_{2}=\sum_{i\in\underline{S}\overline{S}}u_i$ and $U_{3}=\sum_{i\in\overline{S}\underline{S}}u_i$. $U_{2}$ and $U_{3}$ are all Binomial distributed variables and $U_{5}$ is equivalent to a Binomial distributed variable. When $N$ is large, the expected number of sensors in $\underline{S}\overline{S}$, $\overline{S}\underline{S}$ and the expected number of groups in $\underline{T}$ are $NP(i\in\underline{S}\overline{S})$, $NP(i\in\overline{S}\underline{S})$ and $GP(u_i=u_{j}|i,j\in\underline{S}\underline{S})P(i,j\in\underline{S}\underline{S})$, respectively. $P(i\in\underline{S}\overline{S})$ and  $P(i\in\overline{S}\underline{S})$ are defined in \eqref{eq:mu_sigma_en}, and $P(i,j\in\underline{S}\underline{S})$ is defined in \eqref{eq:LL}. $P(u_i=u_{j}|i,j\in\underline{S}\underline{S})$ is given as
\begin{subequations}
    \begin{align}
        P(u_i=u_{j}|i,j\in\underline{S}\underline{S})=&\sum_{q=0,1}P(\mathcal{H}_q)\sum_{t=0,1}P(u_i=t|i\in\underline{S}\underline{S},\mathcal{H}_q)\notag\\
        &P(u_{j}=t|j\in\underline{S}\underline{S},\mathcal{H}_q)\\
        =&(\underline{\pi}_{11}^2+(1-\underline{\pi}_{11})^2)\pi_1\notag\\
        &+(\underline{\pi}_{10}^2+(1-\underline{\pi}_{10})^2)\pi_0
    \end{align}
\end{subequations}
Hence, $U$, which is the sum of Binomial distributed variables, can be approximated as the Gaussian distribution with parameters as follows:
\begin{subequations}
    \begin{align}
        \mu_0^{(RA)}=&E[U|\mathcal{H}_0]\notag\\
        =&GP(u_i=u_{j}|i,j\in\underline{S}\underline{S})P(i,j\in\underline{S}\underline{S})\pi_{10,5}W_{5}\notag\\
        &+N(P(i\in\overline{S}\underline{S})\pi_{10,3}W_{3}+P(i\in\underline{S}\overline{S})\pi_{10,2}W_{2})\\
        \mu_1^{(RA)}=&E[U|\mathcal{H}_1]\notag\\
        &=GP(u_i=u_{j}|i,j\in\underline{S}\underline{S})P(i,j\in\underline{S}\underline{S})\pi_{11,5}W_{5}\notag\\
        &+N(P(i\in\overline{S}\underline{S})\pi_{11,3}W_{3}+P(i\in\underline{S}\overline{S})\pi_{11,2}W_{2})\\
        (\sigma_0^2)^{(RA)}=&Var[U|\mathcal{H}_0]\notag\\
        =&GP(u_i=u_{j}|i,j\in\underline{S}\underline{S})P(i,j\in\underline{S}\underline{S})\notag\pi_{10,5}\notag\\
        &(1-\pi_{10,5})W_{5}^2+N(P(i\in\overline{S}\underline{S})\pi_{10,3}(1-\pi_{10,3})W_{3}^2\notag\\
        &+P(i\in\underline{S}\overline{S})\pi_{10,2}(1-\pi_{10,2})W_{2}^2)\\
        (\sigma_1^2)^{(RA)}=&Var[U|\mathcal{H}_0]\notag\\
        =&GP(u_i=u_{j}|i,j\in\underline{S}\underline{S})P(i,j\in\underline{S}\underline{S})\pi_{11,5}\notag\\
        &(1-\pi_{11,5})W_{5}^2+N(P(i\in\overline{S}\underline{S})\pi_{11,3}(1-\pi_{11,3})W_{3}^2\notag\\
        &+P(i\in\underline{S}\overline{S})\pi_{11,2}(1-\pi_{11,2})W_{2}^2)
    \end{align}
\end{subequations}
The threshold $\eta$ for large $N$ is given as
\begin{equation}\label{eq:eta_RA}
\begin{split}
    \eta^{(RA)}=&\log(\frac{\pi_0}{\pi_1})+E(N_{re}^{LL})\log(\frac{1-\pi_{10,5}}{1-\pi_{11,5}})\\
    &+E(N_{re}^{L})\log(\frac{1-\pi_{10,2}}{1-\pi_{11,2}})+E(N_{re}^{U})\log(\frac{1-\pi_{10,3}}{1-\pi_{11,3}}),
\end{split}
\end{equation}
where $E(N_{re}^{L})=NP(i\in\underline{S}\overline{S})$, $E(N_{re}^{U})=NP(i\in\overline{S}\underline{S})$ and $E(N_{re}^{LL})=GP(u_i=u_{j}|i,j\in\underline{S}\underline{S})$. Thus, the probability of error $P_e^{(RA)}$ for the system is expressed as
\begin{equation}
    P_e^{(RA)}=\pi_0Q\left(\gamma_f^{(RA)}\right)+\pi_1Q\left(\gamma_m^{(RA)}\right),
\end{equation}
where $\gamma_f^{(RA)}=\frac{\eta^{(RA)}-\mu_0^{(RA)}}{\sigma_0^{(RA)}}$ and $\gamma_m^{(RA)}=\frac{\mu_1^{(RA)}-\eta^{(RA)}}{\sigma_1^{(RA)}}$ is the argument of function $Q(.)$ for the probability of false alarm and the argument of function $Q(.)$ for the probability of miss detection for the new proposed fusion rule. Fig. \ref{fig:gamma_alp_0.15_0} shows how argument $\gamma_f^{(RA)}$ changes with $p_1$ given specific $p_2$ and $\alpha_0$ when $N=100$, $P_d=0.9$ and $P_f=0.1$. We can observe that the argument $\gamma_f^{(RA)}$ of RAS is larger than that of EAS under intelligent attacks. Since the argument $\gamma_m^{(RA)}$ has similar properties, we only include the simulation results of $\gamma_f^{(RA)}$ in the paper. Note that the larger arguments mean better detection performance. We can observe that our proposed RAS has a significant improvement on the detection performance of the system when $\alpha_0$ is small. Even though the detection performance of the proposed scheme gets close to EAS when $\alpha_0$ approaches 0.5 and $p_1$ is large, the proposed RAS still outperforms EAS and the direct scheme. This improvement becomes more prominent when $p_1$ is relatively small. Moreover, a large $p_1$ can lead to the easier identification of Byzantine nodes. In this case, the FC has the history of all the local decisions it received in the past. And some reputation-based schemes can help the FC to identify the Byzantine nodes\cite{rawat2010collaborative}\cite{tsitsiklis1988decentralized}.
\subsection{The network with multiple clusters}
\begin{figure}[htbp]
    \centering
    {\includegraphics[width=12em,height=10em]{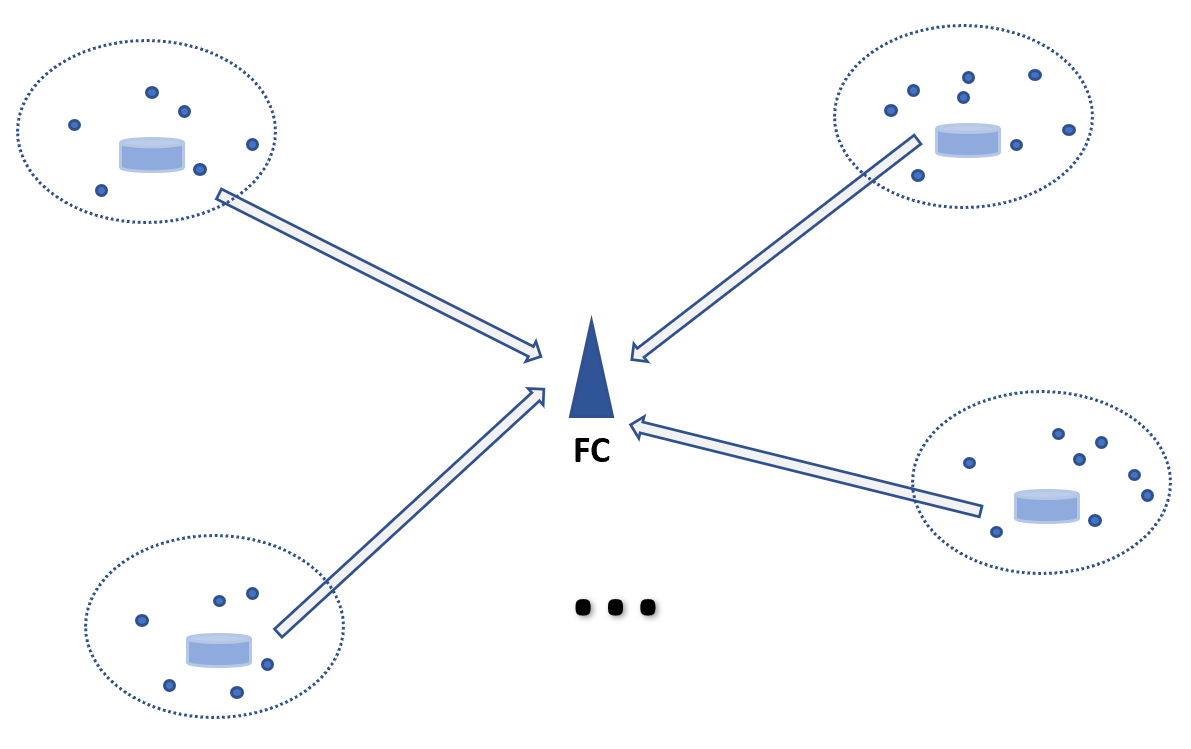}}
    \caption{System model of a distributed CWSN. The blue cylinders represent MMSDs in each cluster and the small blue circles represent low-cost sensors.}
    \label{system_model}
\end{figure}

In this subsection, we extend our work from the single cluster case to the case of multiple clusters in the wide-area network. We show that the proposed RAS can not only improve the detection performance of the system, but also reduce the communication overhead between the clusters and the FC. In a cluster based network as shown in Fig. \ref{system_model}, the $N$ sensors in the network are grouped into $T$ clusters and the sensors in each cluster are further divided into groups of two. Each cluster is equipped with one MMSD which serves as a data integration processor for this cluster. Note that the MMSD is no longer a part of the FC.

Based on the local observations, each sensor makes a binary decision regarding the absence or presence of the PoI. Then, the sensors send both their own decisions and their group member's decision to the corresponding MMSDs. By comparing the MMS of the direct and indirect decisions, the MMSDs are able to obtain the status indicators for all the sensors in the corresponding clusters. Based on these status indicators, each MMSD partitions the sensors in the cluster into sets $\underline{S}\underline{S}$, $\underline{S}\overline{S}$, $\overline{S}\underline{S}$ and $\overline{S}\overline{S}$. In addition, the sensors are placed into $\underline{\mathcal{M}}$ if the local decisions of the sensors in the same group are the same. 

Let $N_t^{(RA)}$ and $N_t^{(A)}$ denote the number of local decisions sent by the MMSDs to the FC for the proposed RAS and the number of local decisions sent by the sensors to the FC, respectively. Note that the MMSDs only transmit the direct decisions, and they do not transmit the ones that satisfy Condition 1 or Condition 2. Thus, the number of direct decisions $N_t^{(RA)}$ sent by the MMSDs to the FC is smaller than that of TAS $N_t^{(A)}$, where $N_t^{(RA)}=|\underline{S}\underline{S}\bigcap\underline{\mathcal{M}}|+|\underline{S}\overline{S}|+|\overline{S}\underline{S}|$ and $N_t^{(A)}=2N$. Let $r$ represent different sets as follows. If $r=00$, it refers to the set $\underline{S}\underline{S}\bigcap\underline{\mathcal{M}}$; If $r=01$ it refers to the set $\underline{S}\overline{S}$; If $r=10$, it refers to the set $\overline{S}\underline{S}$. Each MMSD sends three data packets which contain $r$ and the direct decisions from the sensors in the sets $\underline{S}\underline{S}\bigcap\underline{\mathcal{M}}$, $\underline{S}\overline{S}$ and $\overline{S}\underline{S}$, respectively. For example, if sensor $1$ to sensor $4$ are in $\underline{S}\underline{S}\bigcap\underline{\mathcal{M}}$, sensor $5$ to sensor $8$ are in $\overline{S}\underline{S}$ and sensor $9$ to sensor $12$ are in $\underline{S}\overline{S}$. The three data packets contain $[r=00,u_1,\dots,u_4]$, $[r=10,u_5,\dots,u_8]$ and $[r=01,u_9,\dots,u_{12}]$. Upon receiving these data packets, the FC is able to determine which sets those sensors belong to so that it can make the final decision based on those transmitted direct decisions.

\begin{figure}[htbp]
	\centerline{\includegraphics[width=18em,height=11em]{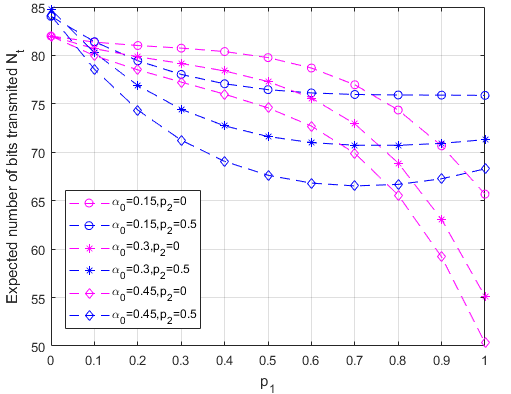}}
	\caption{The expected number of bits transmitted to the FC $N_t$ versus $p_1$ given different value of $\alpha_0$ and $p_2$.}
	\label{fig:Nt}
\end{figure}

When $N$ is large, we are able to calculate the expected number of bits transmitted to the FC from all the MMSDs, which is $E(N_t^{(RA)})=E(N_{re}^{LL})+E(N_{re}^{L})+E(N_{re}^{U})$, according to \eqref{eq:eta_RA}. Fig. \ref{fig:Nt} shows the expected number of bits transmitted to the FC when $N=100$ and $N_t^{(A)}=2N=200$. We can observe that the expected number of bits transmitted to the FC for the proposed RAS significantly decreases compared with the one for TAS. It is due to fact that the MMSDs only send the direct decisions of sensors which do not satisfy Condition 1 or Condition2. We can also observed that the expected number of bits decreases with an increased $\alpha_0$ given a specific $p_2$. It is due to the fact that the number of sensors temporarily removed by the MMSDs increases when the fraction of Byzantine nodes $\alpha_0$ increases with a given attacking probability $p_2$. Hence, the proposed new fusion rule is able to reduce the energy cost of the sensors to half of the traditional case which prolongs the lifetime of the network, especially for the wide area network.

\section{CONCLUSION}
In this work, an audit based mechanism is utilized to mitigate the effect of Byzantine attacks in the networks. Instead of employing the identical attacking strategy of TAS where each sensor utilizes the same attacking probability to falsify the decisions coming from their group member and its own decision, we considered intelligent attackers that can use different attacking strategies. We showed the that it was possible for the intelligent attackers to blind the FC as far as the information conveyed by the audit bits in TAS is concerned. To overcome this problem, we proposed an enhanced audit bit based scheme, namely EAS. Our results showed that the proposed scheme outperforms TAS. Furthermore, we proposed a reduced audit bit based scheme (RAS) based on our new proposed EAS. We showed that RAS is able to further improve the robustness and the detection performance of the system. We extended our work for the wide-area CWSNs. In wide-area cluster-based WSNs, we showed that the proposed RAS is able to significantly reduce the communication overhead between the clusters and the FC. In the future, we intend to consider the scenarios where the CHs (or MMSDs) could also be compromised.

\appendices
\section{Proof of Theorem 1}\label{Pf:P_e}
Instead of directly analyzing the property of $P_e^I$ in terms of $p_2$, we utilize Bhattacharyya distance $\mathcal{BD}$ as a surrogate to asymptotically characterize the detection performance of the system for simplicity. The relationship between Bhattacharyya distance and the probability of error $P_e^I$ is $\lim_{N\rightarrow\infty}\frac{ln(P_e^I)}{N}\leq \mathcal{BD}$. For discrete probability distribution, $\mathcal{BD}=\sum_{\mathbf{u}\in\mathcal{U}}-ln\sqrt{P(\mathbf{u}|\mathcal{H}_1)P(\mathbf{u}|\mathcal{H}_0)}$, where $\mathcal{U}=\{\mathbf{u}_1,\mathbf{u}_2,\dots,\mathbf{u}_{2^N}\}$ is the set of all the possible realizations of vector $\mathbf{u}=[u_1,u_2,\dots,u_N]$. Let $f_i(u_i|i\in\underline{S})=P(u_i|\mathcal{H}_1,i\in\underline{S})P(u_i|\mathcal{H}_0,i\in\underline{S})$ and $f_i(u_i|i\in\overline{S})=P(u_i|\mathcal{H}_1,i\in\overline{S})P(u_i|\mathcal{H}_0,i\in\overline{S})$. Due to the fact that sensors independently send their local decisions, $\mathcal{BD}$ is given as
\begin{equation}
\begin{split}
\mathcal{BD}&=\sum_{\mathbf{u}\in\mathcal{U}}-ln\sqrt{\prod_{i\in\underline{S}}f_i(u_i|i\in\underline{S})\prod_{i\in\overline{S}}f_i(u_i|i\in\overline{S}})\\
&=\sum_{\mathbf{u}\in\mathcal{U}}-ln\sqrt{\prod_{i=1}^N\mathcal{F}_i(u_i)}\\
&=\sum_{\mathbf{u}\in\mathcal{U}}-ln\sqrt{\prod_{i=1}^N\left(\sum_{d_i\in\mathcal{Q}}\mathcal{F}_i(u_i|d_i)P(d_i)\right)}\\
&=\sum_{\mathbf{u}\in\mathcal{U}}-ln\sqrt{\prod_{i=1}^NE_{d_i}\{\mathcal{F}_i(u_i|d_i)\}}
\end{split}
\end{equation}
where $\mathcal{Q}=\{0,1\}$, $\mathbf{d}=[d_1,d_2,\dots,d_N]$ and $d_i\in\mathcal{Q}$. $\mathcal{F}_i(u_i|d_i)=(\overline{\pi}_{11}^{u_i}(1-\overline{\pi}_{11})^{1-u_i}\overline{\pi}_{10}^{u_i}(1-\overline{\pi}_{10})^{1-u_i})^{1-d_i}(\underline{\pi}_{11}^{u_i}(1-\underline{\pi}_{11})^{1-u_i}\underline{\pi}_{10}^{u_i}(1-\underline{\pi}_{10})^{1-u_i})^{d_i}$. $d_i=1$ indicates that the sensor $i$ is placed in Set $\underline{S}$, otherwise, it is placed in Set $\overline{S}$. For sensor $i$, $E_{d_i}\{\mathcal{F}(u_i|d_i)\}$ is given as
\begin{equation}
\begin{split}
&E_{d_i}\{\mathcal{F}(u_i|d_i)\}\\
&\quad\quad=\sum_{q=0,1}\mathcal{F}(u_i|d_i=q)P(d_i=q)\\
&\quad\quad=\overline{\pi}_{11}^{u_i}(1-\overline{\pi}_{11})^{1-u_i}\overline{\pi}_{10}^{u_i}(1-\overline{\pi}_{10})^{1-u_i}P(d_i=1)\\
&\quad\quad\quad+\underline{\pi}_{11}^{u_i}(1-\underline{\pi}_{11})^{1-u_i}\underline{\pi}_{10}^{u_i}(1-\underline{\pi}_{10})^{1-u_i}P(d_i=0).
\end{split}
\end{equation}

 We now have following two cases:
\paragraph{$u_i=1$} In this case, $E_{d_i}\{\mathcal{F}(u_i|d_i)\}=\overline{\pi}_{11}\overline{\pi}_{10}P(d_i=1)+\underline{\pi}_{11}\underline{\pi}_{10}P(d_i=0)$. We know that $P(d_i=1)+P(d_i=0)=1$ and $\underline{\alpha}^I\leq\alpha_0\leq\overline{\alpha}^I$. Let $h(t)=\pi_{11}\pi_{10}$ where $t=\alpha p_1$ is the random variable here.
We can obtain $\frac{\partial^2 h(t)}{t^2}=2(1-2P_d)(1-2P_f)<0$. Hence, $h(t)$ is a concave function and has the property as following.
\begin{equation}\label{eq:proty}
\begin{split}
    	&P(d_i=1)h(t_1)+P(d_i=0)h(t_2)\\
    	&\leq h(P(d_i=1)t_1+P(d_i=0)t_2)=h(t_0)
\end{split}
\end{equation}
where $t_1=\overline{\alpha}^Ip_1$, $t_2=\underline{\alpha}^Ip_1$ and $t_0=\alpha_0p_1$.
\paragraph{$u_i=0$}In this case, $E_{d_i}\{\mathcal{F}(u_i|d_i)\}=(1-\overline{\pi}_{11})(1-\overline{\pi}_{10})P(d_i=1)+(1-\underline{\pi}_{11})(1-\underline{\pi}_{10})P(d_i=0)$. Let $g(t)=(1-\pi_{11})(1-\pi_{10})$ where $t=\alpha p_1$ is the random variable here.
We can obtain $\frac{\partial^2 g(t)}{t^2}=2(1-2P_d)(1-2P_f)<0$. Hence, $g(t)$ is also a concave function and follows the similar property as \eqref{eq:proty}.

Note that we have $\underline{\pi}_{11}=\overline{\pi}_{11}=\pi_{11}$ and $\underline{\pi}_{10}=\overline{\pi}_{10}=\pi_{10}$ when $p_2=0$ according to Lemma 1. We can conclude that $E_{d_i}\{\mathcal{F}(u_i|d_i)\}\leq \mathcal{F}^0(u_i)$, where $\mathcal{F}^0(u_i)=\pi_{11}^{u_i}(1-\pi_{11})^{1-u_i}\pi_{10}^{u_i}(1-\pi_{10})^{1-u_i}$. We call the grouping in TAS with $p_2=0$ as non-effective grouping which is the same as the direct scheme, i.e., $\underline{\alpha}^I=\alpha_0=\overline{\alpha}^I$, and the grouping in TAS with $p_2\neq0$ as effective grouping. According to \eqref{eq:proty}, We show that the Bhattacharyya distance of the effective grouping is always larger than that of the non-effective grouping. According to the analysis above, the detection error $P_e^{(I)}$ can achieve the maximum value when $p_2=0$ given specific $\alpha_0$, $P_d$, $P_f$ and $p_1$. The probability of error for the system with direct scheme is
 \begin{equation}\label{eq:pe_d}
     P_e^{(D)}=\pi_0Q\left(\gamma_f^{(D)}\right)+\pi_1Q\left(\gamma_m^{(D)}\right),
 \end{equation}
 where $\gamma_f^{(D)}$ and $\gamma_m^{(D)}$ are expressed, respectively, as
\begin{subequations}\label{eq:gamma_D}
\begin{align}
    \gamma_f^{(D)}&=Q\left(\frac{\log(\frac{\pi_0}{\pi_1})/\sqrt{N}+\sqrt{N}D_0(\alpha_0,p)}{\sqrt{\pi_{10}(1-\pi_{10})W_d^2}} \right)\\
    \gamma_m^{(D)}&=Q\left(\frac{\log(\frac{\pi_0}{\pi_1})/\sqrt{N}+\sqrt{N}D_1(\alpha_0,p)}{\sqrt{\pi_{11}(1-\pi_{11})W_d^2}} \right),
\end{align}
\end{subequations}
and, $D_0(\alpha_0,p)=\pi_{10}\log(\frac{\pi_{10}}{\pi_{11}})+(1-\pi_{10})\log(\frac{1-\pi_{10}}{1-\pi_{11}})$, $D_1(\alpha_0,p)=\pi_{11}\log(\frac{\pi_{11}}{\pi_{10}})+(1-\pi_{11})\log(\frac{1-\pi_{11}}{1-\pi_{10}})$ and $W_d=\log(\frac{\pi_{11}(1-\pi_{10})}{\pi_{10}(1-\pi_{11})})$. Thus, for the non-effective grouping, according to \eqref{eq:gamma_D}, $D_0(\alpha_0,p)=0$ can make the system be totally blind when $N$ is large enough. We can easily obtain that $D_0(\alpha_0,p)=0$ when $\alpha_0p=\frac{1}{2}$. 

\bibliography{refer.bib}
\bibliographystyle{IEEEtran}
\end{document}